\documentclass[prb,aps,twocolumn,nopacs,superscriptaddress,nofootinbib]{revtex4}
\usepackage{graphicx}
\usepackage{dcolumn}
\usepackage{bm}
\usepackage{amsmath}
\usepackage{epsfig}
\usepackage{color}

\def\ii{{\rm i}}  \def\ee{{\rm e}}
  \def\Rb{{\bf R}}    
      
    \def\kparb{{\bf k}_\parallel}
    \def\gb{{\bf g}}
\begin{document}
\title{Graphene: Free electron scattering within an inverted honeycomb lattice}

\author{Z. M. Abd El-Fattah}
\email[Corresponding author: ]{z.m.abdelfattah@azhar.edu.eg, Zakaria.Eldegwy@icfo.eu}
\affiliation{ICFO-Institut de Ciencies Fotoniques, The Barcelona Institute of Science and Technology, 08860 Castelldefels (Barcelona), Spain}
\affiliation{Physics Department, Faculty of Science, Al-Azhar University, Nasr City, E-11884, Cairo, Egypt}
\author{  M. A. Kher-Elden}
\affiliation{Physics Department, Faculty of Science, Al-Azhar University, Nasr City, E-11884, Cairo, Egypt}
\author{I. Piquero-Zulaica}
\affiliation{Centro de F\'\i sica de Materiales (CSIC-UPV-EHU) and
Materials Physics Center (MPC), 20018 San Sebasti\'an, Spain}
\author{F. J. Garc\'{\i}a de Abajo} 
\email[Corresponding author: ]{Javier.GarciaDeAbajo@icfo.eu}
\affiliation{ICFO-Institut de Ciencies Fotoniques, The Barcelona Institute of Science and Technology, 08860 Castelldefels (Barcelona), Spain}
\affiliation{ICREA-Instituci\'o Catalana de Recerca i Estudis Avan\c{c}ats, Passeig Llu\'{\i}s Companys 23, 08010 Barcelona, Spain}
\author{J. E. Ortega}
\email[Corresponding author: ]{enrique.ortega@ehu.eus}
\affiliation{Centro de F\'\i sica de Materiales (CSIC-UPV-EHU) and
Materials Physics Center (MPC), 20018 San Sebasti\'an, Spain}
\affiliation{Donostia International Physics Center, Paseo Manuel
Lardiz\'abal 4, E-20018 Donostia-San Sebasti\'an, Spain}
\affiliation{Departamento de F\'{\i}sica Aplicada I,
Universidad del Pa\'{\i}s Vasco, 20018 San Sebasti\'an, Spain}

\date{\today}

\begin{abstract}

Theoretical progress in graphene physics has largely relied on the application of a simple nearest-neighbor tight-binding model capable of predicting many of the electronic properties of this material. However, important features that include electron-hole asymmetry and the detailed electronic bands of basic graphene nanostructures (e.g., nanoribbons with different edge terminations) are beyond the capability of such simple model. Here we show that a similarly simple plane-wave solution for the one-electron states of an atom-based two-dimensional potential landscape, defined by a single fitting parameter (the scattering potential), performs better than the standard tight-binding model, and levels to density-functional theory in correctly reproducing the detailed band structure of a variety of graphene nanostructures. In particular, our approach identifies the three hierarchies of nonmetallic armchair nanoribbons, as well as the doubly-degenerate flat bands of free-standing zigzag nanoribbons with their energy splitting produced by symmetry breaking. The present simple plane-wave approach holds great potential for gaining insight into the electronic states and the electro-optical properties of graphene nanostructures and other  two-dimensional materials with intact or gapped Dirac-like dispersions.

\end{abstract}

%\pacs{73.20.At, 73.22.-f, 79.60.Jv}

\maketitle

The two-dimensional (2D) honeycomb carbon-atom lattice known as graphene \cite{GN07} is a promising material for applications in optical and electronic devices \cite{LYU11,VVC12, FS10}. In particular, its peculiar conical electronic dispersion \cite{GWS1984,CGP09} and 2D character enable a uniquely large optical tunability \cite{FRA12,paper196} and a suitable  playground for quantum electrodynamics phenomena, such as the relativistic  Klein tunneling \cite{MIK06}, as well as a customizable zoo of exotic band structures when decorated with defects \cite{FZW18}, arranged in twisted bilayers \cite{YC18}, or laterally patterned into ribbons \cite{MF96, KN96}. Energy-gap engineering in  graphene, an essential prerequisite for nanoelectronics  applications, demands controlled and selective sub-lattice perturbations at the atomic scale, such as chemical doping \cite{JTL15,GRS08} or gating \cite{P10_3}, lateral strain \cite{NYL08,CWN17}, and substrate-induced sublattice asymmetry \cite{ZGF07,VBH14, PRS12,VMS12}.

Graphene nanoribbons  (GNRs) have been extensively studied as simple, appealing nanostructures that lead to electronic band features, such as gap opening, due to quantum confinement, and peculiar edge states that can readily be tuned  through their width, shape, and edge-terminations \cite{MF96, KN96}. The rapidly progressing  on-surface chemistry, which allows  controlled-synthesis of  novel graphene-based nanostructures,  such as GNRs  with complex architectures \cite{CCC15,DDU18,BDR18,GNR_junctions,GNR_junctions2,MVK18},  combined with the precise mapping of their electronic structures using angle-resolved photomission spectroscopy (ARPES)  and scanning tunneling spectroscopy (STS) \cite{RCP12,SUF18,CGC18,MJH14}, make GNRs promising candidates for the  realization of exotic graphene-based nanodevices  \cite{IGM10,LFB17, CNT16}.

Theoretical understanding and prediction of extended  graphene and GNRs properties has been instrumental in the development of the field. Density-functional theory (DFT) accurately describes  their electronic structures, but simpler methods are preferred because they allow us to gain further physical insight. In particular,  following the pioneering work of Wallace \cite{W1947}, the tight-binding (TB) model has  played a central role in the theoretical description of the electronic structure of extended graphene,  yielding remarkable agreement with  DFT calculations. 
 However, noticeable discrepancies between TB and DFT show up when describing GNRs  with either armchair (AGNR) or zigzag (ZGNR) edge terminations. For example, the widely used nearest-neighbors TB  predicts two families of AGNRs, namely semiconductor and metallic, depending on  the number of carbon-dimer lines along the ribbon width ($N_a$)  \cite{KN96}, while  three semiconductor categories are obtained from DFT calculations \cite{SCL06_2} in agreement with STS experiments \cite{KED15,MGC17,STG15,LCW14}.

Both nearest-neighbors TB and nearly-free electron (NFE) models are well-known textbook approaches for  band-structure calculations in solids \cite{K1987,AM1976}. Within the NFE framework, plane wave expansions (PWEs) of the electron states have traditionally played an important role, for example in the description of  electron scattering in metallic and molecular superlattices \cite{paper030,comm001,PLS17}. In particular,  2D hexagonal superlattices, which are known to exhibit graphene-like band structures with  $\overline{ M}$-point gap and  symmetry-protected degeneracy at the $\overline{ K}$-points \cite{paper170,paper175}, are well described  by the PWE approach. Unfortunately, such simple PWEs have not been used for the description of extended graphene or GNRs, although a close correspondence between the TB and NFE models was demonstrated for the so-called $\it{molecular}$ graphene, in which the  Shockley surface state confined by  a hexagonal CO superlattice was shown to exhibit a Dirac-like dispersion \cite{GMK12}.

Here, we demonstrate that the electronic characteristics of  \emph{atomic} graphene  could be finely reproduced via a simple NFE model with a single fitting parameter, namely the scattering potential. In this context, the graphene non-Bravais honeycomb lattice is alternatively modeled as a 2D hexagonal lattice made of the 6-fold symmetric, hexagonally-warped inner part of the carbon rings, where a sufficiently large repulsive potential ($V_3$) is assigned [Fig. 1 (a)]. The potential barrier $V_3$ in reality delimits the attractive Coulomb potential of each carbon atom ($V_1$ and $V_2$).  Perfect agreement with DFT calculations is obtained for the  band structure,  local density of states (LDOSs), and  constant energy surfaces (CESs) using an Electron-Plane-Wave-Expansion (EPWE)  implementation (see Methods).  Interestingly, with the same single fitting parameter $V_3$, the model captures the three categories of AGNRs in decent agreement with DFT. Likewise,  the  1D-bulk band structure and the nearly-degenerate edge state for ZGNRs are obtained in agreement with TB and DFT without any assistance of  electron-electron interactions. Additionally, we find that when the symmetry of the  two carbon sublattices is broken for ZGNRs ($V_1 \neq V_2$), which is a common situation for graphene grown onto different substrates,  the edge state of ZGNRs is split in energy, here without the incorporation of electron-electron interactions. We believe that this simplified picture can be efficiently applied  to explore different varieties of atomic graphene-like extended and finite structures.

\begin{figure}[tb!]
\includegraphics[width=85mm, angle=0,clip]{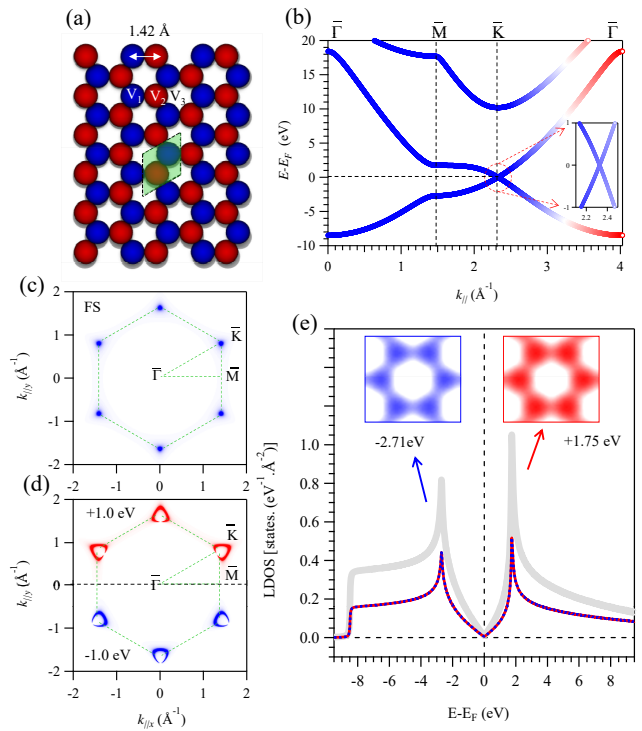}
\caption[Free-standing graphene]{\textbf {Free-standing graphene:} (a) Geometry and scattering potential used in our EPWE calculations. The red and blue circles represent  the two carbon sublattices  with inner potentials $V_1$ and $V_2$, while white regions define the hexagonally-warped bases with an inner potential $V_3$.  (b) Calculated band structure using EPWE along the $\overline{\Gamma M K \Gamma}$ direction, for $V_3$=23 eV and $V_1$=$V_2$=0. The inset is a close up view at the $\overline{ K}$-point. (c-d) Simulated CES taken at  (c) 0 eV and at (d) +1 eV (top) and -1 eV (bottom). (e)  LDOS calculated at the center of the  carbon atoms in each sublattice (red and blue) and  total DOS per unit cell (gray). The insets show the 2D LDOS  at the borders of the $\overline{ M}$-point gap.}
\label{fig1}
\end{figure}

Figure 1 summarizes the electronic characteristics of free-standing graphene, as determined within the EPWE approach. The potential landscape used in the calculation is depicted in Fig. 1(a). The red and blue circles define the position of the carbon atoms, each of radius  $a/2=0.71\,\AA$, and the white regions stand for the carbon-free voids. The unit cell, enclosing one void and two carbon atoms, is marked by the black lines (shaded area), with unit-cell vectors of length $d=\sqrt{3}\times1.42\,\AA$. The band structure presented in Fig. 1(b) is obtained by setting the potential difference between the voids ($V_3$ = 23 eV) and the two equivalent carbon atoms ($V_1$ = $V_2$ = 0) to  $\Delta$V$ $ = 23 eV and the effective mass ($m_\text{{eff}}$ ) to unity. These parameters, which coincide with the binding energy of the carbon 2$s$ level  \cite{carbon_2s} and the free electron mass ($m_\text{{e}}$), nicely reproduce the band structure of free-standing graphene obtained from DFT calculations and experiments \cite{PL15,RRF13}. The Fermi energy ($E_F$) is set at the non-gapped (see zoom-in) Dirac point, and the  $\overline{\Gamma}$-point energy is accordingly found at $\sim$ -8.5 eV. The lower and upper edges of the  $\overline{M}$-point gap are  -2.71 eV and +1.75 eV, respectively, while the slope of the linear bands at the $\overline{K}$ points is $\simeq6.5\,$eV\,$\AA$ [i.e., Fermi velocity ($v_F$) $\simeq$ 1 $\times$ 10$^{6}$ m/s], in perfect agreement with literature values \cite{DV15}. We stress that small deviations from the employed value of $\Delta$$V$ and  $m_\text{{eff}}$  yield noticeable changes on the relative energetic position of the bands,  the size of the $\overline{M}$-point gap, and the degree of electron-hole asymmetry [see Supplementary Information (SI), Fig. S1).

In Fig 1(c-d) we present the simulated photoemission intensity of the constant energy surfaces (CESs). The Fermi surface (FS) consists of single spots centered at the six $\overline{K}$-points of the Brillouin zone (BZ, green lines), resulting from intact Dirac cones [Fig. 1(c)]. At lower and higher binding energies ($e.g.,$ +1 eV and -1 eV), these spots diverge into the characteristic graphene triangular pockets, as shown, respectively, in the upper (red) and lower (blue) panels  of Fig. 1(d). The trigonal shape of these CESs further reassures  that the electron-hole asymmetry present in (b), which in TB calculations is accounted for by employing additional hopping parameters for second/third nearest neighbors $t'$/$t''$ \cite{CGP09,KYJ13,RMT02}, is  naturally captured by EPWE. Indeed, the different hopping parameters employed in TB  are consistent with the different effective potentials ($e.g.,$ height $\times$ width) felt by electrons moving from one carbon towards  neighboring atoms. Furthermore, the variation of the photoemission intensity within the trigonal pockets agrees nicely  with recent ARPES experiments \cite{FVH14,ULO18,VSS08,VBH14}. The electronic characteristics of graphene, as deduced from the total-DOS (TDOS)  and LDOS, are also presented in Fig. 1(e). The V-like peculiarities at $E_F$ are revealed in the TDOS per unit cell (gray) and the LDOS at the two carbon atoms (blue and red),  all exhibiting clear electron-hole asymmetry. The onsets of LDOS at $\sim$ -8.5 eV define the $\overline{\Gamma}$-point energy, whereas the peaks at -2.71 eV and +1.75 eV are the borders of the $\overline{M}$-point gap. The LDOS at the two carbon atoms are clearly coincident, a common finding for pristine graphene. The 2D LDOS maps depicted at the insets and taken at the boundaries of the $\overline{M}$-point gap are identical, confirming the absence of $\overline{K}$-point gaps based on symmetry considerations \cite{paper170,paper175}. Given the calculations and analysis presented in Fig. 1, the electronic features of a \emph{free-standing} graphene sheet obtained from experiments and DFT calculations are well-reproduced. 

\begin{figure}[tb!]
  \includegraphics[width=87mm,angle=0,clip]{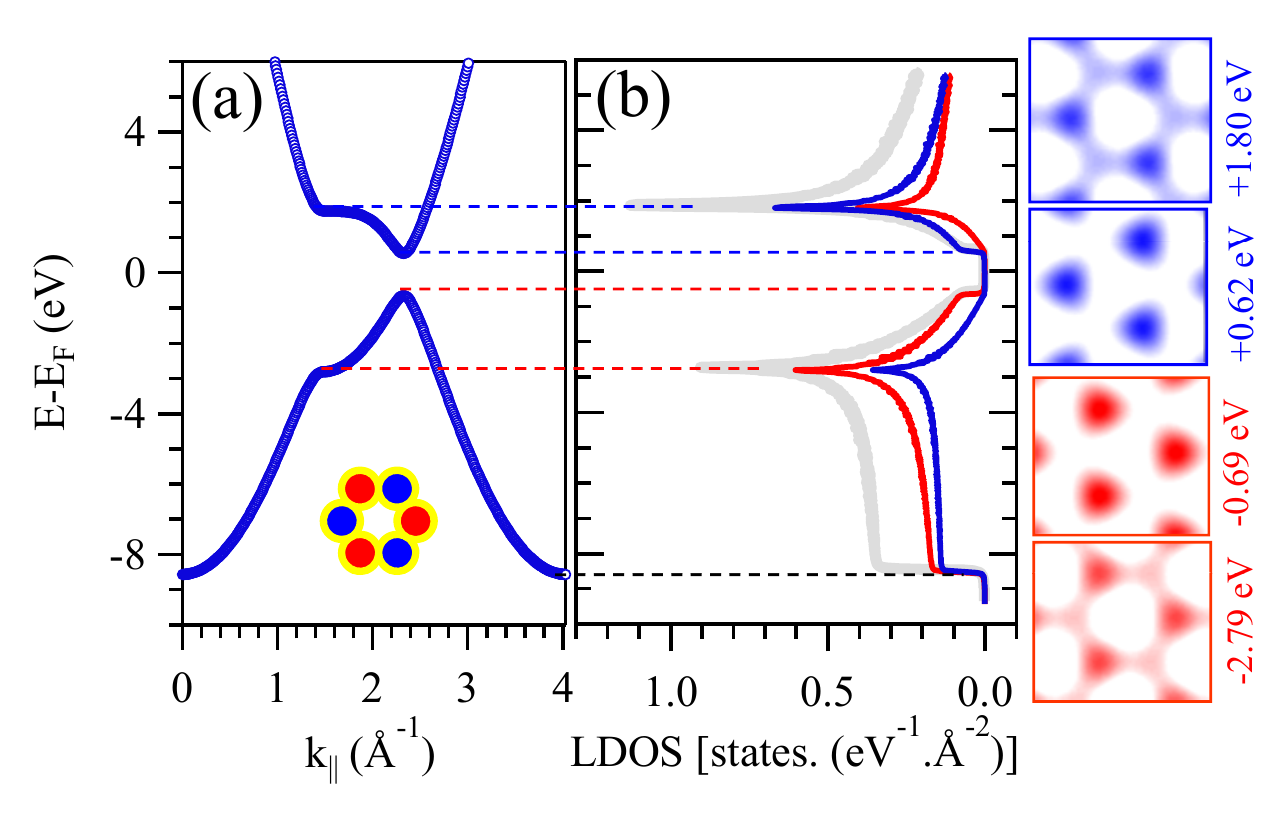}
  \caption[Perturbed graphene]{\textbf {Perturbed graphene:} (a) Calculated band structure using EPWE along the $\overline{\Gamma M K \Gamma}$ direction, for $V_3$=23 eV,  $V_1$=+1  eV, and $V_2$=-1 eV.  (b) The corresponding TDOS (gray) and LDOS taken at the red and blue circles in (a). The right panel in (b) presents the 2D LDOS maps taken at the lower (red) and upper (blue) edges of the $\overline{ M}$-point and $\overline{K}$-point gaps.}
  \label{fig2}
\end{figure}

\begin{figure*}[tb!]
  \includegraphics[width=160mm,angle=0,clip]{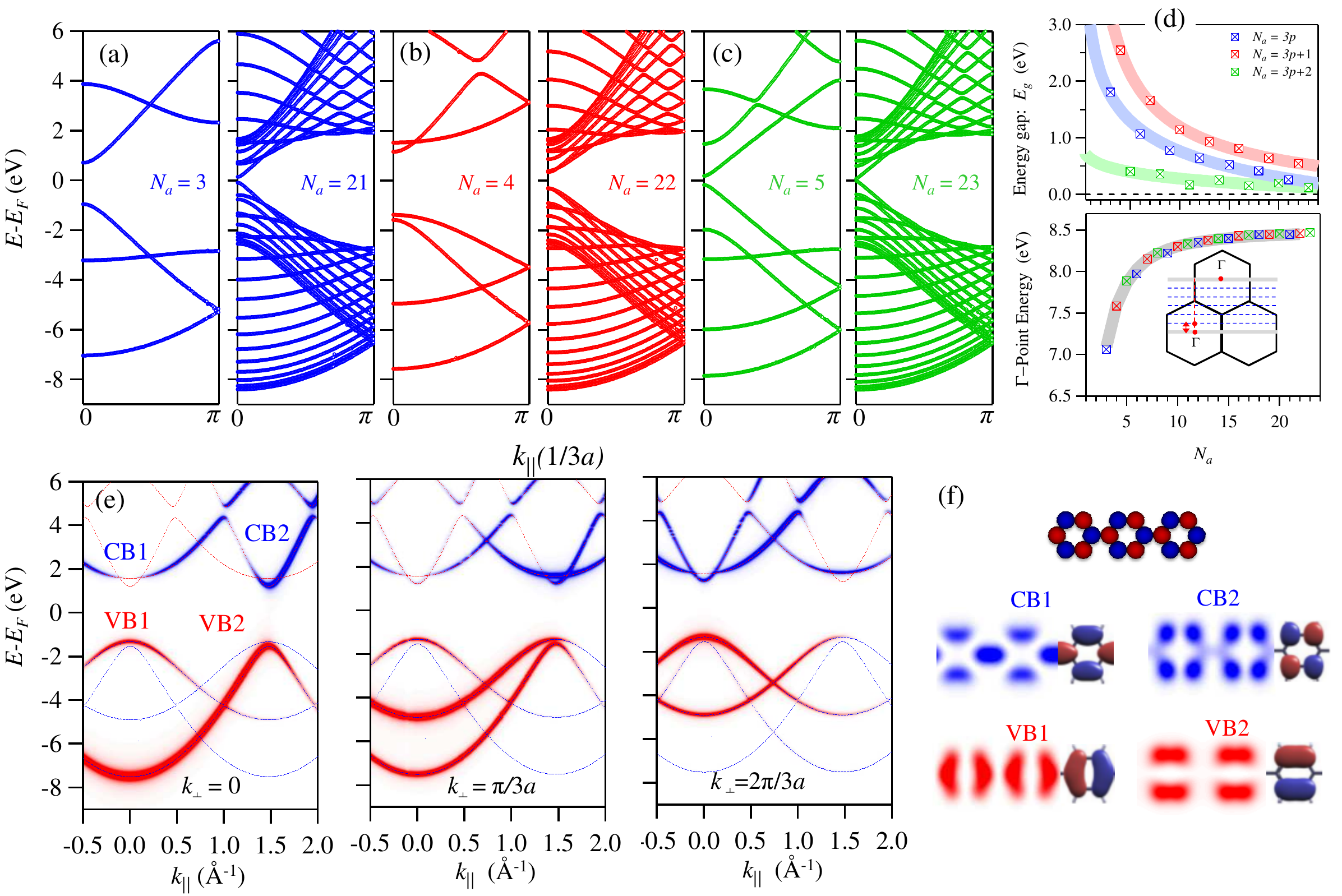}
  \caption{\textbf{Armchair graphene nanoribbons (AGNRs):} (a-c) Band structure  of $N_a$-AGNR  with $N_a$ = 3 and 21 (a), 4 and 22 (b), and 5 and 23 (c), as obtained using EPWE. (d) Variation of the gap size and $\overline{\Gamma}$-point energy for the three categories of AGNRs. The inset shows the first and second BZ of graphene. (e) Simulated photoemission intensity for the 4-AGNR along the ribbon axis taken at $k_{\bot}$ = 0 (left), $k_{\bot}$ = $\pi$/3$a$ (center), and $k_{\bot}$ = $2\pi$/3$a$ (right).(f) LDOS for the 3-AGNR taken at VB1, VB2, CB1, and CB2. The corresponding molecular orbitals deduced from DFT calculations (adapted from Ref.\ \onlinecite{DFT_orbitals}) are shown to the right, and the geometry used in EPWE calculations is shown on top.}
\label{fig3}
\end{figure*}

Perturbations induced by a graphene support (i.e., a substrate, and/or the deposition of adsorbates/dopants) have shown to change the electronic properties of graphene in different ways. Figure 2 presents possible electronic modifications in one of  such  perturbation cases, as calculated using the present EPWE model. In Fig. 2(a), we explore the effect of   broken symmetry for the two carbon atoms on the electronic structure. The band structure is obtained  by assigning different potentials for each carbon-sublattice ($V_1$(red) = -1 eV and  $V_2$(blue) = +1 eV). The main modification is the opening of an energy gap at the $\overline{K}$ point ($E$$_g$($\overline{K}$) = 1.3 eV), which is a natural consequence of the broken symmetry of the potential landscape within the unit cell. Such broken-symmetry-induced gaps have been reported experimentally for different graphene systems, such as  graphene grown on Ir(111) \cite{KPP11}, Ru(0001) \cite{BGW09}, hydrogenated-graphene \cite{BJN10}, and other systems \cite{ZGF07,VBH14,PRS12,VMS12,VSS08}. The large symmetry-induced gap shown here is meant only to highlight the effect, but the actual values of  $V_1$ and $V_2$ can be tuned to yield the experimental gap. We also note that the $\overline{\Gamma}$-point energy is unaltered in this particular example, as the average of the potentials $V_1$ and $V_2$ is zero, and is only shifted in energy otherwise.  The TDOS and LDOS spectra presented in (b) precisely follow these band structure modifications. In addition to the peaks at the borders of the $\overline{M}$-point gap, a deviation from the V-shape peculiarities occurs for all spectra where, instead,  the TDOS and LDOS vanish at the energy range  spanning the $\overline{K}$-point gap boundaries. Particularly relevant is the non-equivalence of the LDOS at the two carbon atoms within the unit cell, where the weight of the LDOS changes from one carbon  (red) to the second sublattice (blue) by crossing the energy gap, which is further shown in the 2D LDOS presented in Fig. 2(b, right). Additional electronic modifications, such as hybridization and doping, which still preserve the $\overline{K}$-point  degeneracy, are briefly presented in SI, Fig. S2.

The calculations/analysis here presented clearly reveal that the electronic structure of graphene could be reproduced by simple geometrical regions in which specific values of the potential are assigned. In what follows, we explore the applicability of this approach to the study of both AGNRs and ZGNRs. We employ the same geometry and potential landscape while varying the ribbon width and termination. We assume that the ribbons are infinitely extended along the ribbon axis ( i.e., the $x$-direction), and  decoupled in $y$-direction by separating them by $\geq$ 20 \AA{} gaps. Figure 3 (a-c) depicts the band structure along the ribbon axis for the three  different classes of AGNRs with  (a) $N_a$ = 3$p$, (b) $N_a$ = 3$p$+1 , and (c) $N_a$ = 3$p$+2 , where $p$ is a positive integer [here, $p$ = 1 (left) and 7 (right)]. In contrast to standard TB  calculations, the three types exhibit energy gaps ($E_g$) with $E_g$(3$p$+1) $>$ $E_g$(3$p$) $>$ $E_g$(3$p$+2), and their width  follows an inverse proportionality (see colored curves in (d)) in agreement with DFT calculations. Notice that  the carbon-carbon distance in the bulk region and at the ribbon boundaries is fixed here to the same value (i.e., structural relaxation is not considered, although DFT  indicates a  $\sim$ 3.5\% contraction in the bond length at the edges) \cite{SCL06_2}, yet the model yields appreciable energy gaps for the 3$p$+2 family. We also show that the $\overline{\Gamma}$-point energy, irrespectively of the ribbon family, asymptotically approaches $\overline{\Gamma}$-point of extended graphene by increasing the width, strictly following the width-dependent $\overline{\Gamma}$ point energy (gray curve) obtained by stepping along the Brillouin zone slices of graphene (see inset to (d)). Particularly relevant is the shallow dispersion of the bands at -2.6 eV for the  $N_a$ = 3 (a) and  $N_a$ = 5 (c), which in standard TB exhibit no dispersion \cite{VFS16,KN96}. This, further, indicates that  the NFE approach naturally considers all possible crosstalk/hopping between neighboring carbon atoms with the scattering potential as a single fitting parameter.

In what follows, we check the capability of our model to simulate the photoemission intensity and density of states for nanoribbons. This should serve as a guidance for experimentalists to perform proper assignments of specific GNR bands, which might become problematic for wider ribbons with nearby dispersing bands, and in general, due to strong variations in intensity caused by effects related to the photoemission matrix element. Figure 3(e) presents the simulated photoemission intensity for the 4-AGNR along the ribbon axis. A subtle variation of photoemission intensity for different bands is clear. For example, at $k_{\bot}$ = 0 (left), the frontier valence band (VB1) has predominantly-symmetric spectral weight  around $k_{||}$ = 0, while VB2  gains spectral weight over wider $k_{||}$  range with asymmetric photoemission intensity around the top of the band ($\sim$ 1.5 \AA{}$^{-1}$). This distribution of photoemission intensity changes drastically at different $k_{\bot}$, such as $k_{\bot}$ = $\pi$/3$a$ (center) and 2$\pi$/3$a$ (right).  Indeed,  these are all 1D bands, and therefore, are not dispersing in the perpendicular direction ($k_{\bot}$), yet strong photoemission intensity modulation  is present for the VB and CB alike [SI, Fig. S3(a)]. Therefore, constant-energy surfaces (i.e., $k_{||}$ vs $k_{\bot}$ maps) such as the one shown in the SI, Fig. S3(b), are essential for a proper assignment of bands. Actually, the simulation of such photoemission intensity maps  has recently solved a long-standing contradiction between STS and ARPES data for the 7-AGNR, where the VB2 was mistakenly assigned in ARPES experiments to the VB1 \cite{SUF18,footnote}. A further confirmation of the full functionality of our model is provided by comparing the calculated 2D-LDOS to the molecular orbitals obtained from DFT calculations, and in particular, for the 3-AGNR, as shown in Fig. 3 (f). The matching between LDOS (left) and DFT orbitals (right) is remarkable \cite{DFT_orbitals}. The overall  agreement  with DFT extends even beyond the description of simple AGNRs: complex graphene-based structures such as zigzag \cite{DFT_orbitals} and heterogeneous ribbons, as well as  nanoporous graphene \cite{MVK18}, are equally well described using our  NFE approach  (see SI, Fig. S4).

\begin{figure}[tb!]
\includegraphics[width=85mm, angle=0,clip]{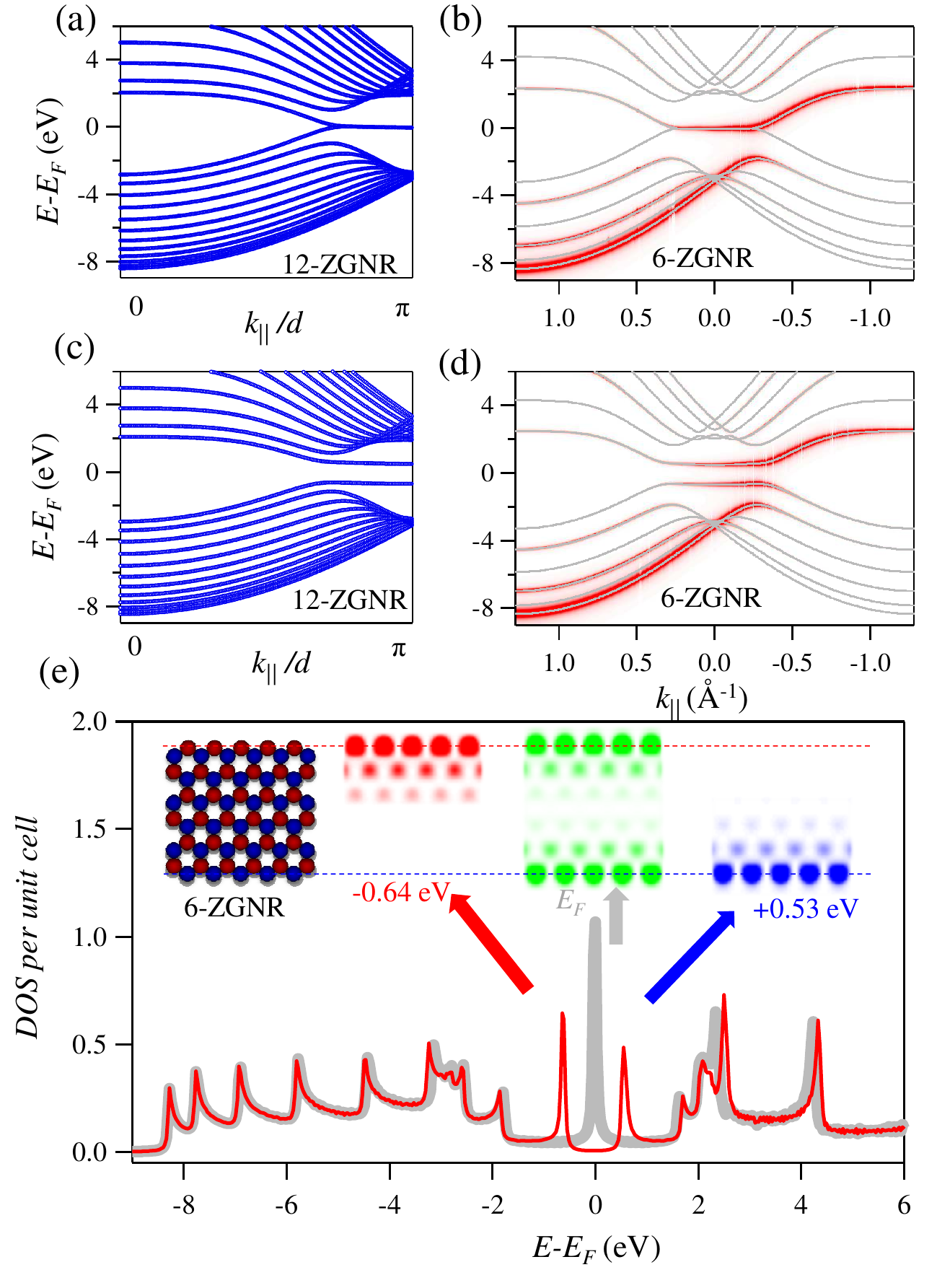}
 \caption{\textbf{Zigzag graphene nanoribbons (ZGNRs):} (a-b) Band structure of 12-ZGNR (a) and  6-ZGNR (b) obtained using EPWE. (c,d) The same as (a,b), but the symmetry of the carbon sublattice is broken. We show the simulated photoemission intensity for the 6-ZGNR along the ribbon taken at $k_{\bot}$ = 0 in (b,d). (e) DOS per unit cell for the 6-ZGNR before (gray) and after (red) breaking the symmetry of the carbon sublattices. The insets show the geometry and 2D-LDOSs at the edge-state energies.}
\label{fig4}
\end{figure}

Likewise, the electronic structure of ZGNRs could be obtained using our EPWE approach. Figure 4(a-b) presents  band structure calculations for selected ZGNRs with $N_a$ = 12 (a) and 6 (b). Their characteristic bulk bands and the energy-degenerate edge states are obtained in agreement with TB and DFT calculations when  on-site Hubbard potentials and exchange interactions, respectively, are not considered. In (b) the simulated photoemission intensity for the 6-ZGNR  is also shown at $k_{\bot}$ = 0, with  the edge state exhibiting a high-enough spectral weight as to be probed by average techniques such as ARPES, provided that ZGNRs aligned on substrates are experimentally available on large areas.

As previously discussed, for  an extended graphene sheet with symmetry-broken carbon sublattices, a $\overline{K}$-point gap opens up (see Fig. 2). Here we demonstrate the effect that this broken symmetry has on the electronic structure of  ZGNRs, specifically on the edge state. By  assigning different values to the potentials at the two carbons sublattices ( $V_1$ = -1 eV and  $V_2$ = +1 eV), taking the 12-ZGNR as an example, we show that the edge state is split in energy, while the 1D-bulk projected bands are practically unaltered [Fig. 4(c)]. The energy gap between the split edge states is the same as the size of the $\overline{K}$-point gap of the corresponding extended graphene presented in Fig. 2(a). Both bands of the energy-split edge state have reasonable  photoemission  spectral weights, as demonstrated for the 6-ZGNR at $k_{\bot}$ = 0 in (d). Finally, we present in Fig. 4(e) the DOS curves for the free-standing (gray) and symmetry-broken (red) sublattices in the 6-ZGNR, where an energy-splitting of the DOS peak at $E_F$ is obtained. Although  the DOS profile for asymmetric ribbons (red) resembles  the one reported experimentally (and from DFT) for the 6-ZGNR \cite{FWY16}, the broken-symmetry-induced gaps could be distinguished from electron-electron-interaction gaps \cite{SCL06_2,SCL06,YCL08} by plotting the 2D spatial distribution of the LDOS taken at the energy of the edge state,  as shown on top of the spectra in  Fig. 4(e). The insets present the EPWE geometry of the 6-ZGNR and the 2D-LDOS taken at the energy of the lower (red) and upper (blue)  edge states, where the LDOS is clearly localized at one edge, while for fully symmetric ribbons both edges are equally occupied at $E_F$ (green). Notice that the one-edge localization of the LDOS has not experimentally been reported, neither for degenerate nor for energy-split edges states,  yet its realization could have potential impact as a switch in 1D conduction-channels through  gating. We also anticipate that the combination of broken symmetry and electron correlation should produce a clear imbalance in the LDOS at both edges of the ribbons, in addition to the intrinsic asymmetry produced by the different dispersion of the upper and lower edges of the gap and the electron-hole asymmetry.

The fact that the extended and finite graphene characteristics are well captured within the framework of the NFE model could have far reaching implications, since some of the electronic structure variations and size dependence are not unique to graphene. Similar atomic systems, such as silicene or boron nitrides, could be understood following the same approach. Furthermore, nanometer-sized ribbons made from hexagonal superlattices, such as molecular graphene \cite{GMK12} or metallic superlattices possessing graphene-like band structures \cite{paper170}, should exhibit these types of size dependence variations. What makes these variations experimentally accessible and potentially relevant for technology in  the graphene case, and in other 2D atomic lattices,  is the combination of a  large $\overline{M}$-point gap (several eV), which  for superlattices reduces to just a few meV, and the steep dispersion near the $\overline{ K}$ points. Finally, this simple NFE description of graphene and its nanostructures should have large impact on the efficient  simulation of graphene-based devices and phenomena, such as negative refraction and super lenses in $p$-$n$ junctions, using, for example, the complementary electronic  boundary-element  method (EBEM) solver, which was previously used to describe similar effects in 2D metallic superlattices \cite{paper153,paper301}. 

In conclusion, we have showed that the electronic structure of the $\pi$ band in free-standing and perturbed graphene can be well described by a simple nearly-free-electron model (EPWE) applied to an inverted honeycomb lattice defined by a sufficiently large confining potential. With the same single fitting parameter (i.e., the scattering barrier) the electronic properties of graphene nanostructures, such as armchair and zigzag GNRs, are well described. Our approach  simplifies the exploration of  newly emerging artificial systems with  fundamental and technological interest, such as nanostructured 2D materials, topological GNR junctions with peculiar end states \cite{GNR_junctions,GNR_junctions2}, and artificial flat band lattices \cite{LAF18}.

\acknowledgments

This work has been supported in part by the Spanish MINECO (Grant Nos. MAT2014-59096-P, MAT-2016-78293-C6, MAT-2017-88374-P, MAT2017-88492-R, and SEV2015-0522), the Basque Government (Grant No. IT-621-13), the Catalan CERCA Program, Fundaci\'o Privada Cellex, and AGAUR (Grant No. 2017 SGR 1651).

\section*{METHODS}

{\bf Effective 2D potential description.} We simulate the electronic structure of graphene in terms of the one-electron states of a 2D potential landscape, in which each carbon atom is represented by a circle filled with uniform potential, embedded in a flat interstitial region (see Fig.\ 1(a)). We then write the Schr\"odinger equation as
\begin{align}
-\frac{\hbar^2}{2m_{\rm eff}}\nabla^2\phi+V\phi=E\phi, \label{eqSchro}
\end{align}
where the energy $E$ is expressed relative to a reference level (e.g., the Dirac point), $m_{\rm eff}$ is the effective mass, $V(\Rb)$ is the 2D potential as a function of spatial coordinates $\Rb=(x,y)$ along the graphene plane, and $\phi(\Rb)$ is the electron wave function.

{\bf Plave-wave expansion for periodic systems.} We take $V$ to be periodic and express it in terms of Fourier components as
\begin{align}
V(\Rb)=\sum_{\gb} V_\gb\,\ee^{\ii\gb\cdot\Rb}, \label{eqV}
\end{align}
where the sum extends over 2D reciprocal lattice vectors $\gb$ with coefficients $V_\gb=(1/A)\int_{\rm 1BZ} d^2\Rb\; V(\Rb)\;\ee^{-\ii\gb\cdot\Rb}$ calculated as an integral over the first Brillouin zone (1BZ), normalized to the unit-cell area $A$. Using Bloch's theorem, we anticipate electron wave functions labeled by a band index $j$ and the 2D wave vector $\kparb$ within the 1BZ:
\begin{align}
\phi_{\kparb j}(\Rb)=\frac{1}{\sqrt{NA}}\sum_{\gb} \phi_{\kparb j,\gb}\,\ee^{\ii(\kparb+\gb)\cdot\Rb}, \label{eqphi}
\end{align}
where we use the (infinite) number of cells $N$ for normalization purposes. Inserting Eqs.\ (\ref{eqV}) and (\ref{eqphi}) into Eq.\ (\ref{eqSchro}), we find the linear system of equations
\begin{align}
&E_{\kparb j}\,\phi_{\kparb j,\gb}= \label{eqperiod}\\
&\sum_{\gb'}\left[(\hbar^2/2m_{\rm eff})\left|\kparb+\gb\right|^2\delta_{\gb,\gb'}+V_{\gb-\gb'}\right]\phi_{\kparb j,\gb'}. \nonumber
\end{align}
Because $V_{\gb-\gb'}$ is a Hermitian matrix with indices $\gb$ and $\gb'$, for each value of $\kparb$ we obtain different bands $j$ of real eigenenergies $E_{\kparb j}$ and eigenstates of coefficients $\phi_{\kparb j,\gb}$. We solve Eq.\ (\ref{eqperiod}) by retaining a finite number of $\gb$'s within a sufficiently large distance $g_{\rm max}$ to the origin in reciprocal space. The eigenstates form an orthonormal system,
\begin{align}
\int d^2\Rb\,\phi_{\kparb j}(\Rb)\phi_{\kparb' j'}^*(\Rb)=\delta_{\kparb\kparb'}\delta_{jj'}, \nonumber
\end{align}
provided we impose the normalization condition $\sum_\gb \left|\phi_{\kparb j,\gb}\right|^2=1$.

{\bf LDOS calculation for periodic systems.} The local density of states (LDOS) is directly calculated from its definition
\begin{align}
{\rm LDOS}(E,\Rb)=\sum_{\kparb j} \left|\phi_{\kparb j}(\Rb)\right|^2\,\delta\left(E-E_{\kparb j}\right). \nonumber
\end{align}
In practice, we use the prescription $\sum_{\kparb} \rightarrow (NA/4\pi^2)\int d^2\kparb$ and evaluate this integral in a dense grid by interpolating the eigenstates and eigenenergies within each grid element.

{\bf Calculation of photoemission angular distributions for periodic systems.} For simplicity, we dismiss the contribution of the normal component of the electron wave function to the photoemission matrix elements, as it should just introduce a smooth and broad angular dependence, which we represent through a multiplicative coefficient $C$ in the resulting photoemission intensity. We focus instead on the contribution of the in-plane wave function and further approximate the parallel component of the photoelectron wave function as a normalized plane wave $\ee^{\ii\kparb^{\rm out}\cdot\Rb}/\sqrt{NA}$. The angle-resolved photoemission intensity corresponding to a binding energy $E$ and photoelectron wave vector $\kparb^{\rm out}$ is then given from Fermi's golden rule as
\begin{align}
\frac{dI(E)}{d^2\kparb^{\rm out}}=&\,C \sum_j\int_{\rm 1BZ}\!d^2\kparb \left|\int\!\!d^2\Rb\,\phi_{\kparb j}(\Rb)\,\frac{\ee^{-\ii\kparb^{\rm out}\cdot\Rb}}{\sqrt{NA}}\right|^2 \nonumber\\
&\quad\quad\quad\times\;\delta\!\left(E-E_{\kparb j}\right) \nonumber\\
=&\,C \sum_j \left|\phi_{\kparb j,\gb}\right|^2 \delta\!\left(E-E_{\kparb j}\right)\bigg|_{\kparb=\kparb^{\rm out}-\gb}, \nonumber
\end{align}
where $\gb$ in the last expression is chosen such that $\kparb$ lies within the 1BZ.

\setcounter{figure}{0}    

\renewcommand{\thefigure}{S\arabic{figure}}%

\section{\textbf{SUPPLEMENTARY INFORMATION}}

\subsection{\textbf{Electron-hole asymmetry}}

The inherent  electron-hole asymmetry of graphene manifests itself as slightly different band dispersions above and below the Dirac-point, and consequently, a clear deviation from  a symmetric V-shape is clearly observable  in the LDOS, as shown in Fig. 1. This asymmetry shows up in conventional  DFT calculations and in experiments, while the popular nearest-neighbors TB model requires additional hopping parameters to account for it \cite{KYJ13}. In contrast, our NFE approach captures  electron-hole asymmetry, which can be  tuned to fit experimental data through a  small adjustment in the effective mass. Figure S1 presents  the band structures (a)  and DOS per unit cell (b) for  $m_\text{eff}$ = (1$\pm$0.05)$m_\text{e}$ compared with  $m_\text{eff}$ = $m_\text{e}$ (in gray). We note the strong modifications of the electronic band structure (about $\pm$ 500 meV variation in the $\overline{\Gamma}$-point energy) with slight (5$\%$) relaxation of $m_\text{eff}$. The inset to (b) highlights the electron-hole asymmetry near the Dirac point \cite{ALD12}.

\begin{figure}[b!]
  \includegraphics[width=90mm,angle=0,clip]{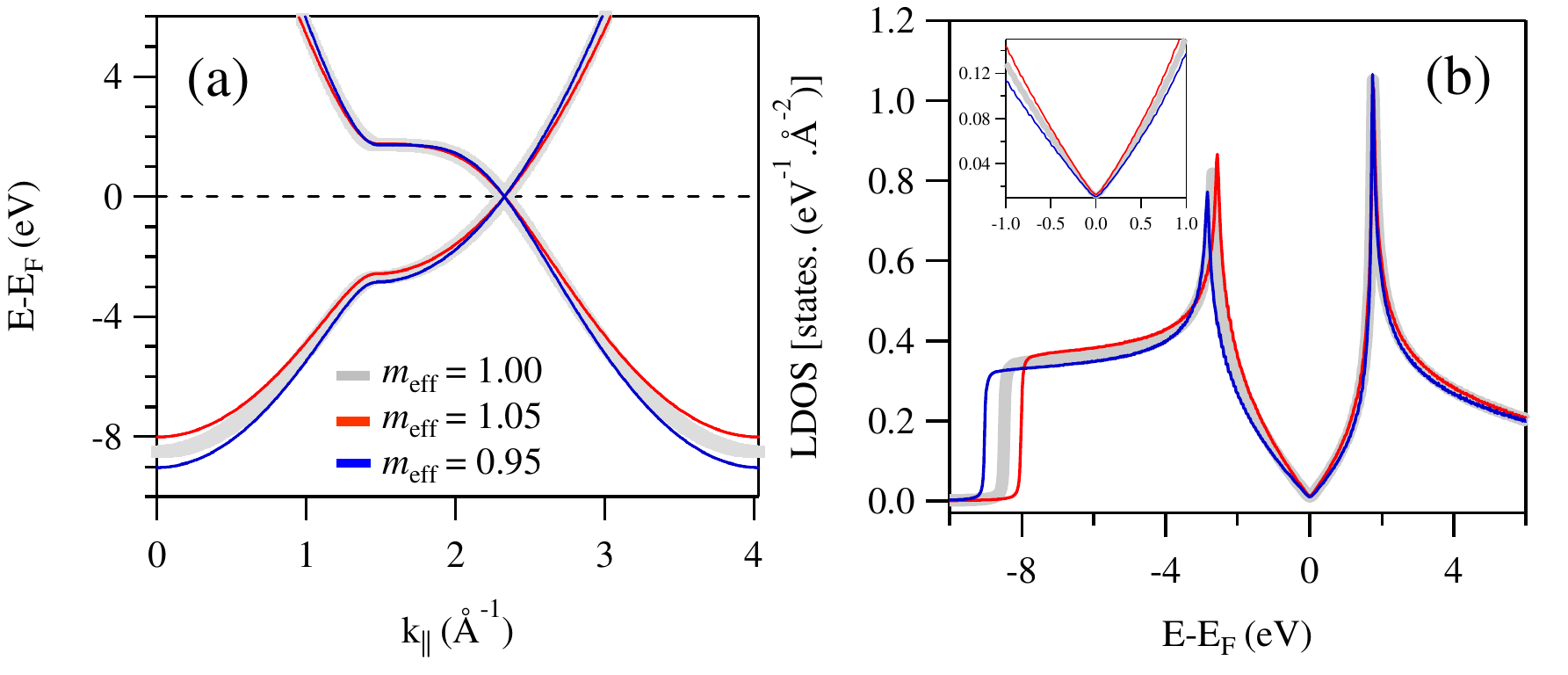}
  \caption[Electron-hole asymmetry]{\textbf {Electron-hole asymmetry:} Calculated band structure along the $\overline{\Gamma M K \Gamma}$ excursion within the first Brillouin zone (a) and DOS per unit cell (b) obtained using EPWE, for $V_3$=23 eV and  $V_1$= $V_2$=0, with $m_\text{{eff}}$ =  1.0 $m_\text{{e}}$  (gray),  1.05 $m_\text{{e}}$  (red), and 0.95 $m_\text{{e}}$  (blue).}
  \label{fig1}
\end{figure}

\subsection{\textbf{Graphene-substrate hybridization and doped graphene}}

It is a well-established result for graphene, or generally for 2D materials with  lattice, that the degeneracy at the $\overline{K}$ points is protected by both structural and time-reversal symmetries. However, for some graphene/substrate combinations, such as graphene on Co(0001), the symmetry of the carbon sublattices is broken while the Dirac cone remains intact. To explain this effect, a dynamical hybridization scenario has been discussed \cite{VMS12}. Such systems can also be modeled within our approach. From a simple NFE model argument, the size of the $\overline{M}$-point gap (and the $\overline{K}$-point gap as well, if allowed by symmetry considerations) scales with the scattering potential. The latter is actually proportional to the product of the barrier height ($V$) and width. By setting the covalent radii of the carbon atoms to  0.63 \AA { }(red) and 0.71 \AA{ }(blue), we can close the $\overline{K}$-point gap, although the symmetry of the voids  is reduced to three-fold, as shown in Fig. S2(a).  In fact, the 2D LDOS map  at the lower and upper edges of the $\overline{M}$-point gap are exactly the same as in $\emph{free-standing}$ graphene [see  Fig. 1(e)], where the LDOS associated with the two carbon atoms are identical. Therefore, it is difficult to distinguish {\it free-standing} and asymmetric non-gapped graphene based on LDOS data, such as provided by STS experiments. However, the $\overline{\Gamma}$-point energy is shifted towards lower binding energy as a result of the increased area of the voids at the expense of the covalent radii of the carbon atoms. For larger radii, where an overlap between the carbon atoms occurs, the $\overline{\Gamma}$-point shifts instead toward higher binding energy. We note that the closing of the $\overline{K}$-point gap in the symmetry-broken graphene of Fig. S2(a) is always associated with $p$ or $n$ doping. This conditional doping  can be used in experiments to distinguish asymmetric non-gapped from pristine graphene.  We also note that the doping here differs from  conventional  $p$- or $n$-doping of  pristine graphene, in which there is no  breaking of the symmetry of the carbon atoms, thus leading to a rigid downward/upward energy shift , as shown in Fig. S2(b). Here, the band structures for $n$-doped (green) and $p$-doped (red) graphene are obtained by setting  $V_1$ = $V_2$ = -1 eV and $V_1$ = $V_2$ = +1 eV, respectively.

\begin{figure}[tb!]
  \includegraphics[width=80mm,angle=0,clip]{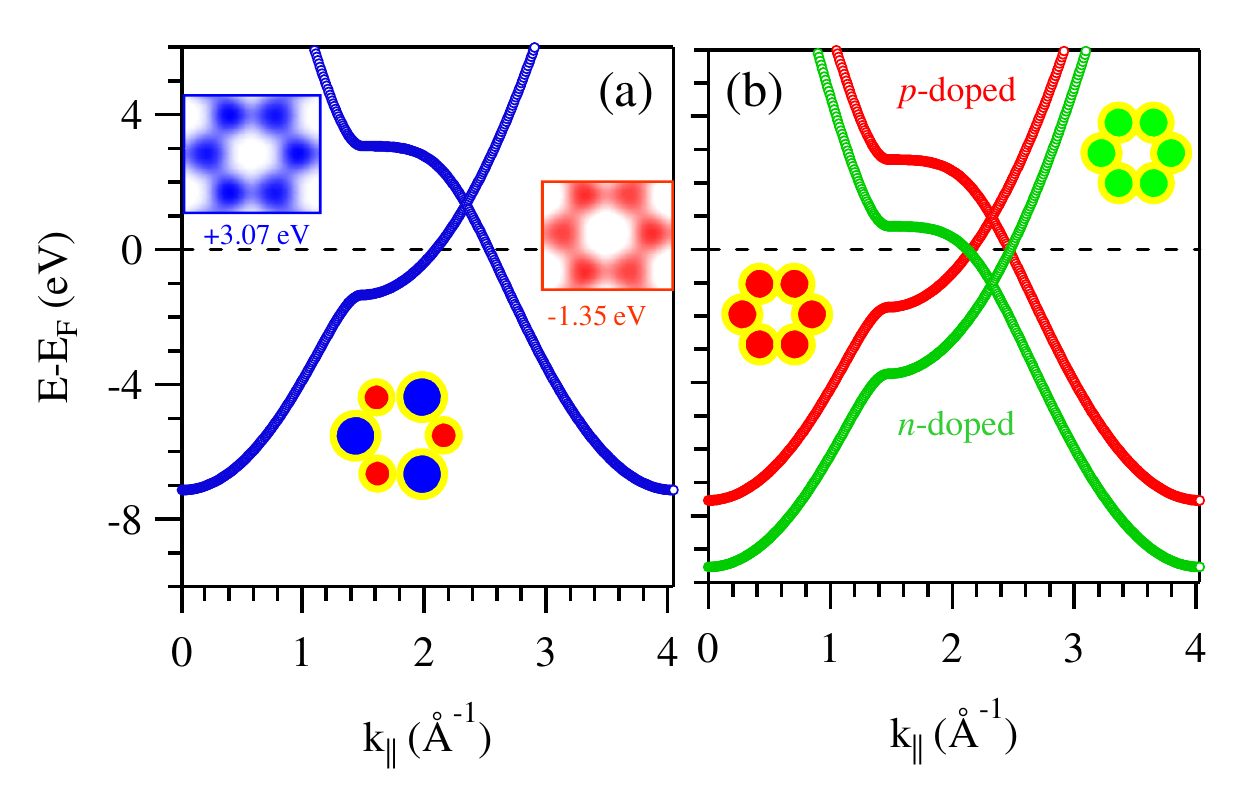}
  \caption[Perturbed graphene]{\textbf {Graphene-substrate hybridization and doped graphene:} (a) Graphene electron band structure calculated  using EPWE along the $\overline{\Gamma M K \Gamma}$ excursion, for $V_3$=23 eV,  $V_1$=+1  eV, and $V_2$=-1 eV,  with different covalent radii assigned to the carbon atoms (blue and red). The insets show 2D LDOS maps taken at the borders of the $\overline{ M}$-point gap. (d) Band structures of $p$-type (red) and $n$-type (green) doped-graphene obtained by setting $V_1$=$V_2$=+1  eV and $V_1$=$V_2$= -1  eV, respectively.}
  \label{fig1}
\end{figure}

\begin{figure*}[tb!]
  \includegraphics[width=150mm,angle=0,clip]{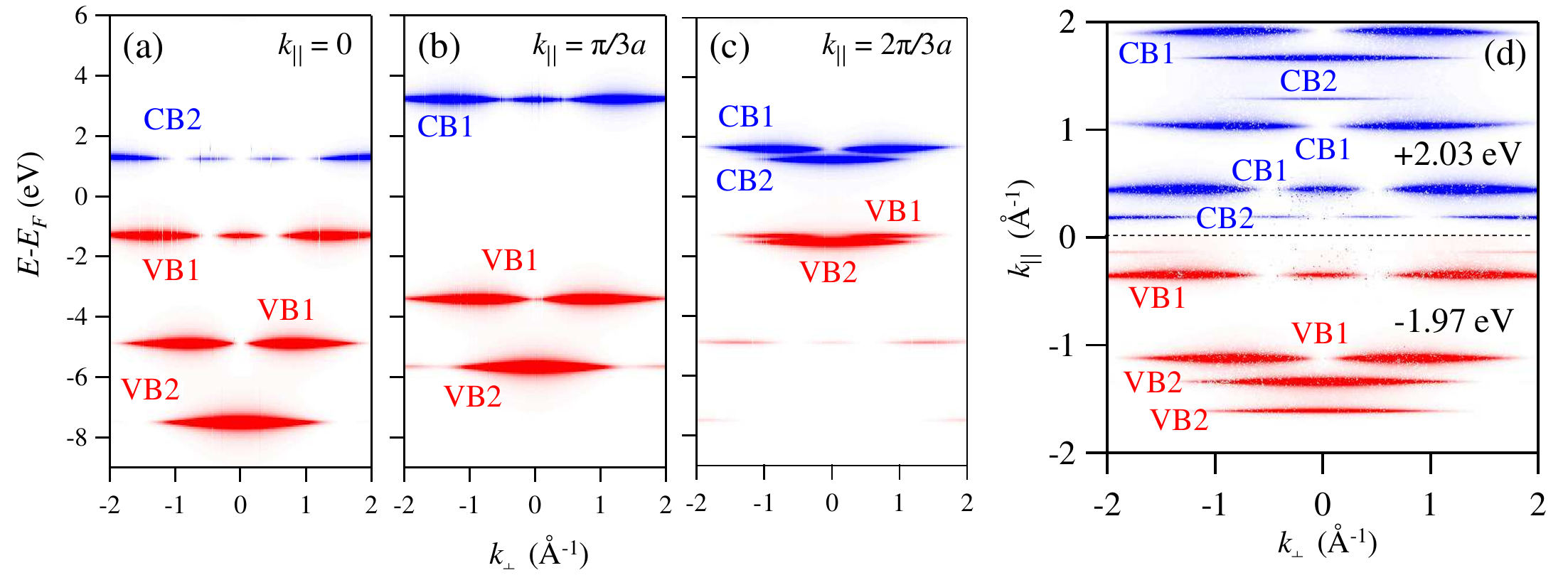}
  \caption[Perturbed graphene]{\textbf { Photoemission intensity:}  Simulated photoemission intensity for the 4-AGNR as a function of electron wave vector component perpendicular to the ribbon axis, taken at $k_{||}$ = 0 (a), $k_{||}$ = $\pi$/3$a$ (b), and $k_{||}$ = $2\pi$/3$a$ (c). (d) Constant-energy surfaces taken close to the bottom of  the CB (top) and at the top of the VB (bottom).}
\label{fig1}
\end{figure*}

\begin{figure*}[tb!]
  \includegraphics[width=130mm,angle=0,clip]{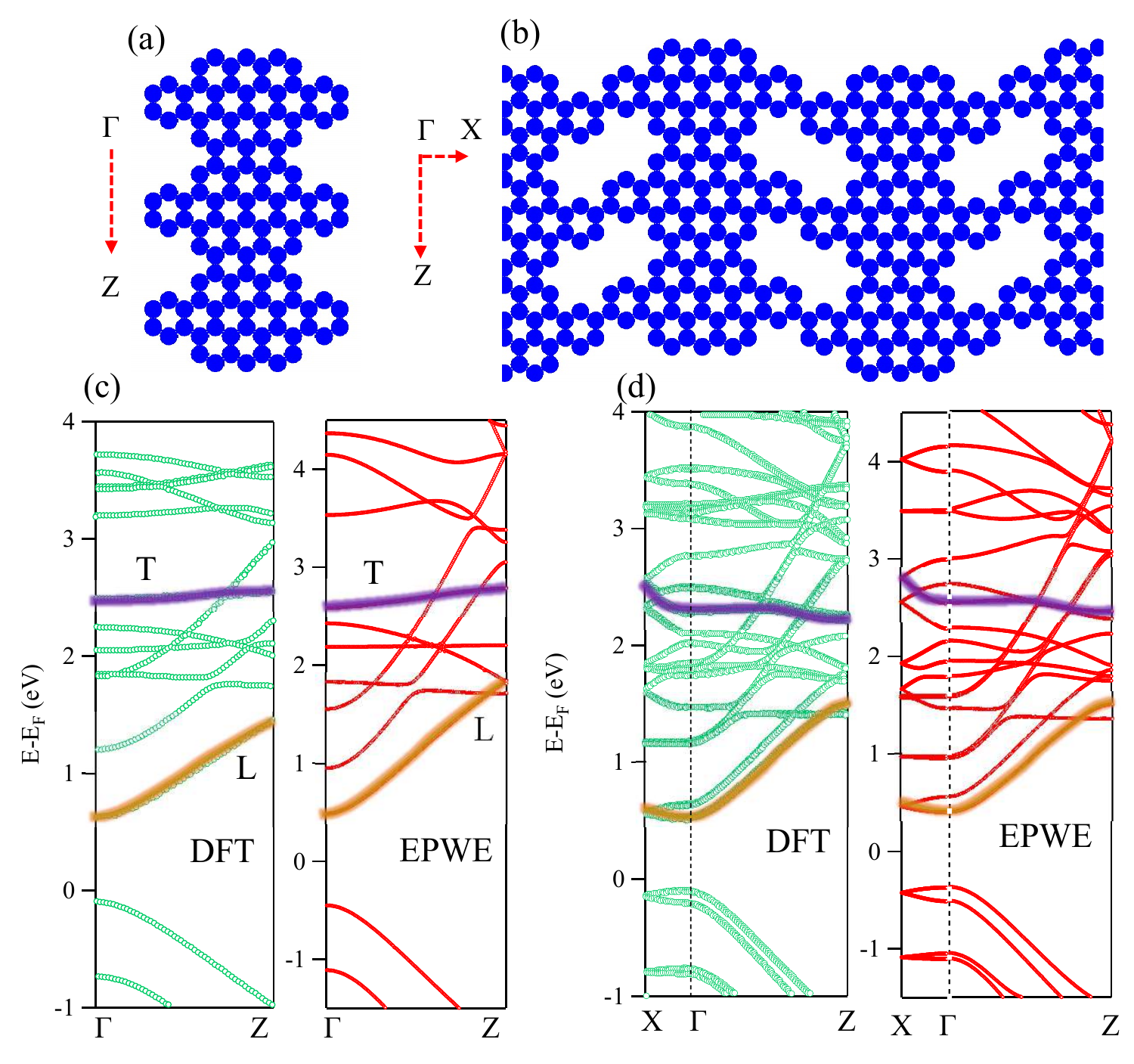}
  \caption[Graphene nanoribbons (GNRs)]{\textbf{Heterogeneous AGNR and nanoporous graphene (NPG):} (a,b) Potential landscape used in EPWE calculations for (a)  7-13-AGNR and (b) NPG.  The corresponding  band structures are shown, respectively, in the right panels of (c) and (d). The DFT-calculated band structures (c,d: left) are also shown  for reference (adapted from Ref.\ \onlinecite{MVK18}).}
  \label{fig3}
\end{figure*}

\subsection{\textbf{Photoemission intensity}}

 In Fig. S3(a-c) we show the photoemission intensity  of the 1D bands (i.e., perpendicular to the ribbon axis) for the 4-AGNR. All bands exhibit a strong intensity modulation, so that specific bands light up at selective values of $k_{||}$. For example, CB2 and CB1 have spectral weight for $k_{||}$ = 0 (a) and $k_{||}$ = $\pi$/3$a$ (b), respectively, while both VB1 and VB2 exhibit peaks  at $k_{||}$ = $2\pi$/3$a$ (c). The  $k_{||}$ vs $k_{\bot}$ maps   taken at 2.03 eV (top) and -1.95 eV (bottom) are presented in (d), allowing us to obtain a proper band  assignment  from ARPES measurements, which have been routinely simulated in recent works \cite{SUF18,DFT_orbitals,VB_B-7AGNR2}.

\subsection{\textbf{Heterogeneous AGNR and nanoporous graphene (NPG)}}

In this section, we demonstrate the ability of our NFE approach to describe the electronic  structure of  more complex graphene-based nanostructures. We take the 7-13-AGNR  structure \cite{MVK18}, which has a modulation in width between 7 and 13 carbon atoms, as an example for heterogeneous nanoribbons and junctions [Fig. S4(a)]. We also consider nanoporous graphene (NPG) \cite{MVK18} made from the connections between these self-assembled  7-13-AGNR in an out-of-phase  configuration [Fig. S4(b)]. In the potential landscapes shown in (a-b), the blue circles define the carbon atoms with $V_1$=$V_2$=0, while the void and pore potentials are set to $V_3$=23 eV. The band structures obtained from our EPWE approach [Fig. S4(c-d), right] are compared with the ones obtained from DFT  [Fig. S4(c-d), left] \cite{MVK18}. Our model clearly captures the essential features of the DFT band structures, including the gap size,  as well as the L and T states.


\begin{thebibliography}{72}
\expandafter\ifx\csname natexlab\endcsname\relax\def\natexlab#1{#1}\fi
\expandafter\ifx\csname bibnamefont\endcsname\relax
  \def\bibnamefont#1{#1}\fi
\expandafter\ifx\csname bibfnamefont\endcsname\relax
  \def\bibfnamefont#1{#1}\fi
\expandafter\ifx\csname citenamefont\endcsname\relax
  \def\citenamefont#1{#1}\fi
\expandafter\ifx\csname url\endcsname\relax
  \def\url#1{\texttt{#1}}\fi
\expandafter\ifx\csname urlprefix\endcsname\relax\def\urlprefix{URL }\fi
\providecommand{\bibinfo}[2]{#2}
\providecommand{\eprint}[2][]{\url{#2}}

\bibitem[{\citenamefont{Geim and Novoselov}(2007)}]{GN07}
\bibinfo{author}{\bibfnamefont{A.~K.} \bibnamefont{Geim}} \bibnamefont{and}
  \bibinfo{author}{\bibfnamefont{K.~S.} \bibnamefont{Novoselov}},
  \bibinfo{journal}{Nat.\ Mater.} \textbf{\bibinfo{volume}{6}},
  \bibinfo{pages}{183} (\bibinfo{year}{2007}).

\bibitem[{\citenamefont{Liu et~al.}(2011)\citenamefont{Liu, Yin, Ulin-Avila,
  Geng, Zentgraf, Ju, Wang, and Zhang}}]{LYU11}
\bibinfo{author}{\bibfnamefont{M.}~\bibnamefont{Liu}},
  \bibinfo{author}{\bibfnamefont{X.}~\bibnamefont{Yin}},
  \bibinfo{author}{\bibfnamefont{E.}~\bibnamefont{Ulin-Avila}},
  \bibinfo{author}{\bibfnamefont{B.}~\bibnamefont{Geng}},
  \bibinfo{author}{\bibfnamefont{T.}~\bibnamefont{Zentgraf}},
  \bibinfo{author}{\bibfnamefont{L.}~\bibnamefont{Ju}},
  \bibinfo{author}{\bibfnamefont{F.}~\bibnamefont{Wang}}, \bibnamefont{and}
  \bibinfo{author}{\bibfnamefont{X.}~\bibnamefont{Zhang}},
  \bibinfo{journal}{Nature} \textbf{\bibinfo{volume}{474}}, \bibinfo{pages}{64}
  (\bibinfo{year}{2011}).

\bibitem[{\citenamefont{Vicarelli et~al.}(2012)\citenamefont{Vicarelli,
  Vitiello, Coquillat, Lombardo, Ferrari, Knap, Polini, Pellegrini, and
  Tredicucci}}]{VVC12}
\bibinfo{author}{\bibfnamefont{L.}~\bibnamefont{Vicarelli}},
  \bibinfo{author}{\bibfnamefont{M.~S.} \bibnamefont{Vitiello}},
  \bibinfo{author}{\bibfnamefont{D.}~\bibnamefont{Coquillat}},
  \bibinfo{author}{\bibfnamefont{A.}~\bibnamefont{Lombardo}},
  \bibinfo{author}{\bibfnamefont{A.~C.} \bibnamefont{Ferrari}},
  \bibinfo{author}{\bibfnamefont{W.}~\bibnamefont{Knap}},
  \bibinfo{author}{\bibfnamefont{M.}~\bibnamefont{Polini}},
  \bibinfo{author}{\bibfnamefont{V.}~\bibnamefont{Pellegrini}},
  \bibnamefont{and}
  \bibinfo{author}{\bibfnamefont{A.}~\bibnamefont{Tredicucci}},
  \bibinfo{journal}{Nat.\ Mater.} \textbf{\bibinfo{volume}{11}},
  \bibinfo{pages}{4176} (\bibinfo{year}{2012}).

\bibitem[{\citenamefont{Schwierz}(2010)}]{FS10}
\bibinfo{author}{\bibfnamefont{F.}~\bibnamefont{Schwierz}},
  \bibinfo{journal}{Nat.\ Nanotech.} \textbf{\bibinfo{volume}{5}},
  \bibinfo{pages}{487} (\bibinfo{year}{2010}).

\bibitem[{\citenamefont{Semenoff}(1984)}]{GWS1984}
\bibinfo{author}{\bibfnamefont{G.~W.} \bibnamefont{Semenoff}},
  \bibinfo{journal}{Phys. Rev. Lett.} \textbf{\bibinfo{volume}{53}},
  \bibinfo{pages}{2449} (\bibinfo{year}{1984}).

\bibitem[{\citenamefont{{Castro Neto} et~al.}(2009)\citenamefont{{Castro Neto},
  Guinea, Peres, Novoselov, and Geim}}]{CGP09}
\bibinfo{author}{\bibfnamefont{A.~H.} \bibnamefont{{Castro Neto}}},
  \bibinfo{author}{\bibfnamefont{F.}~\bibnamefont{Guinea}},
  \bibinfo{author}{\bibfnamefont{N.~M.~R.} \bibnamefont{Peres}},
  \bibinfo{author}{\bibfnamefont{K.~S.} \bibnamefont{Novoselov}},
  \bibnamefont{and} \bibinfo{author}{\bibfnamefont{A.~K.} \bibnamefont{Geim}},
  \bibinfo{journal}{Rev.\ Mod.\ Phys.} \textbf{\bibinfo{volume}{81}},
  \bibinfo{pages}{109} (\bibinfo{year}{2009}).

\bibitem[{\citenamefont{Fei et~al.}(2012)\citenamefont{Fei, Rodin, Andreev,
  Bao, McLeod, Wagner, Zhang, Zhao, Thiemens, Dominguez et~al.}}]{FRA12}
\bibinfo{author}{\bibfnamefont{Z.}~\bibnamefont{Fei}},
  \bibinfo{author}{\bibfnamefont{A.~S.} \bibnamefont{Rodin}},
  \bibinfo{author}{\bibfnamefont{G.~O.} \bibnamefont{Andreev}},
  \bibinfo{author}{\bibfnamefont{W.}~\bibnamefont{Bao}},
  \bibinfo{author}{\bibfnamefont{A.~S.} \bibnamefont{McLeod}},
  \bibinfo{author}{\bibfnamefont{M.}~\bibnamefont{Wagner}},
  \bibinfo{author}{\bibfnamefont{L.~M.} \bibnamefont{Zhang}},
  \bibinfo{author}{\bibfnamefont{Z.}~\bibnamefont{Zhao}},
  \bibinfo{author}{\bibfnamefont{M.}~\bibnamefont{Thiemens}},
  \bibinfo{author}{\bibfnamefont{G.}~\bibnamefont{Dominguez}},
  \bibnamefont{et~al.}, \bibinfo{journal}{Nature}
  \textbf{\bibinfo{volume}{487}}, \bibinfo{pages}{82} (\bibinfo{year}{2012}).

\bibitem[{\citenamefont{Chen et~al.}(2012)\citenamefont{Chen, Badioli,
  Alonso-Gonz\'alez, Thongrattanasiri, Huth, Osmond, Spasenovi\'c, Centeno,
  Pesquera, Godignon et~al.}}]{paper196}
\bibinfo{author}{\bibfnamefont{J.}~\bibnamefont{Chen}},
  \bibinfo{author}{\bibfnamefont{M.}~\bibnamefont{Badioli}},
  \bibinfo{author}{\bibfnamefont{P.}~\bibnamefont{Alonso-Gonz\'alez}},
  \bibinfo{author}{\bibfnamefont{S.}~\bibnamefont{Thongrattanasiri}},
  \bibinfo{author}{\bibfnamefont{F.}~\bibnamefont{Huth}},
  \bibinfo{author}{\bibfnamefont{J.}~\bibnamefont{Osmond}},
  \bibinfo{author}{\bibfnamefont{M.}~\bibnamefont{Spasenovi\'c}},
  \bibinfo{author}{\bibfnamefont{A.}~\bibnamefont{Centeno}},
  \bibinfo{author}{\bibfnamefont{A.}~\bibnamefont{Pesquera}},
  \bibinfo{author}{\bibfnamefont{P.}~\bibnamefont{Godignon}},
  \bibnamefont{et~al.}, \bibinfo{journal}{Nature}
  \textbf{\bibinfo{volume}{487}}, \bibinfo{pages}{77} (\bibinfo{year}{2012}).

\bibitem[{\citenamefont{Katselson et~al.}(2006)\citenamefont{Katselson,
  Novoselov, and Geim}}]{MIK06}
\bibinfo{author}{\bibfnamefont{M.~I.} \bibnamefont{Katselson}},
  \bibinfo{author}{\bibfnamefont{K.~S.} \bibnamefont{Novoselov}},
  \bibnamefont{and} \bibinfo{author}{\bibfnamefont{A.~K.} \bibnamefont{Geim}},
  \bibinfo{journal}{Nat.\ Phys.} \textbf{\bibinfo{volume}{2}},
  \bibinfo{pages}{620} (\bibinfo{year}{2006}).

\bibitem[{\citenamefont{Forsythe et~al.}(2018)\citenamefont{Forsythe, Zhou,
  Watanabe, Taniguchi, Pasupathy, Moon, Koshino, Kim, and Dean}}]{FZW18}
\bibinfo{author}{\bibfnamefont{C.}~\bibnamefont{Forsythe}},
  \bibinfo{author}{\bibfnamefont{X.}~\bibnamefont{Zhou}},
  \bibinfo{author}{\bibfnamefont{K.}~\bibnamefont{Watanabe}},
  \bibinfo{author}{\bibfnamefont{T.}~\bibnamefont{Taniguchi}},
  \bibinfo{author}{\bibfnamefont{A.}~\bibnamefont{Pasupathy}},
  \bibinfo{author}{\bibfnamefont{P.}~\bibnamefont{Moon}},
  \bibinfo{author}{\bibfnamefont{M.}~\bibnamefont{Koshino}},
  \bibinfo{author}{\bibfnamefont{P.}~\bibnamefont{Kim}}, \bibnamefont{and}
  \bibinfo{author}{\bibfnamefont{C.~R.} \bibnamefont{Dean}},
  \bibinfo{journal}{Nat.\ Nanotech.} \textbf{\bibinfo{volume}{13}},
  \bibinfo{pages}{566} (\bibinfo{year}{2018}).

\bibitem[{\citenamefont{Cao et~al.}(2018)\citenamefont{Cao, Fatemi, Fang,
  Watanabe, Taniguchi, Kaxiras, and Jarillo-Herrero}}]{YC18}
\bibinfo{author}{\bibfnamefont{Y.}~\bibnamefont{Cao}},
  \bibinfo{author}{\bibfnamefont{V.}~\bibnamefont{Fatemi}},
  \bibinfo{author}{\bibfnamefont{S.}~\bibnamefont{Fang}},
  \bibinfo{author}{\bibfnamefont{K.}~\bibnamefont{Watanabe}},
  \bibinfo{author}{\bibfnamefont{T.}~\bibnamefont{Taniguchi}},
  \bibinfo{author}{\bibfnamefont{E.}~\bibnamefont{Kaxiras}}, \bibnamefont{and}
  \bibinfo{author}{\bibfnamefont{P.}~\bibnamefont{Jarillo-Herrero}},
  \bibinfo{journal}{Nature} \textbf{\bibinfo{volume}{556}}, \bibinfo{pages}{43}
  (\bibinfo{year}{2018}).

\bibitem[{\citenamefont{Fujita et~al.}(1996)\citenamefont{Fujita, Wakabayashi,
  Nakada, and Kusakabe}}]{MF96}
\bibinfo{author}{\bibfnamefont{M.}~\bibnamefont{Fujita}},
  \bibinfo{author}{\bibfnamefont{K.}~\bibnamefont{Wakabayashi}},
  \bibinfo{author}{\bibfnamefont{K.}~\bibnamefont{Nakada}}, \bibnamefont{and}
  \bibinfo{author}{\bibfnamefont{K.}~\bibnamefont{Kusakabe}},
  \bibinfo{journal}{J.\ Phys.\ Soc.\ Jpn.} \textbf{\bibinfo{volume}{65}},
  \bibinfo{pages}{1920} (\bibinfo{year}{1996}).

\bibitem[{\citenamefont{Nakada et~al.}(1996)\citenamefont{Nakada, Fujita,
  Dresselhaus, and Dresselhaus}}]{KN96}
\bibinfo{author}{\bibfnamefont{K.}~\bibnamefont{Nakada}},
  \bibinfo{author}{\bibfnamefont{M.}~\bibnamefont{Fujita}},
  \bibinfo{author}{\bibfnamefont{G.}~\bibnamefont{Dresselhaus}},
  \bibnamefont{and} \bibinfo{author}{\bibfnamefont{M.~S.}
  \bibnamefont{Dresselhaus}}, \bibinfo{journal}{Phys.\ Rev.\ B}
  \textbf{\bibinfo{volume}{54}}, \bibinfo{pages}{17954} (\bibinfo{year}{1996}).

\bibitem[{\citenamefont{Joucken et~al.}(2015)\citenamefont{Joucken, Tison,
  F\`evre, Tejeda, Taleb-Ibrahimi, E.~Conrad, Chacon, Bellec, Girard, Rousset
  et~al.}}]{JTL15}
\bibinfo{author}{\bibfnamefont{F.}~\bibnamefont{Joucken}},
  \bibinfo{author}{\bibfnamefont{Y.}~\bibnamefont{Tison}},
  \bibinfo{author}{\bibfnamefont{P.~L.} \bibnamefont{F\`evre}},
  \bibinfo{author}{\bibfnamefont{A.}~\bibnamefont{Tejeda}},
  \bibinfo{author}{\bibfnamefont{A.}~\bibnamefont{Taleb-Ibrahimi}},
  \bibinfo{author}{\bibfnamefont{V.~R.} \bibnamefont{E.~Conrad}},
  \bibinfo{author}{\bibfnamefont{C.}~\bibnamefont{Chacon}},
  \bibinfo{author}{\bibfnamefont{A.}~\bibnamefont{Bellec}},
  \bibinfo{author}{\bibfnamefont{Y.}~\bibnamefont{Girard}},
  \bibinfo{author}{\bibfnamefont{S.}~\bibnamefont{Rousset}},
  \bibnamefont{et~al.}, \bibinfo{journal}{Sci.\ Rep.}
  \textbf{\bibinfo{volume}{5}}, \bibinfo{pages}{14564} (\bibinfo{year}{2015}).

\bibitem[{\citenamefont{Gierz et~al.}(2008)\citenamefont{Gierz, Riedl, Starke,
  Ast, and Kern}}]{GRS08}
\bibinfo{author}{\bibfnamefont{I.}~\bibnamefont{Gierz}},
  \bibinfo{author}{\bibfnamefont{C.}~\bibnamefont{Riedl}},
  \bibinfo{author}{\bibfnamefont{U.}~\bibnamefont{Starke}},
  \bibinfo{author}{\bibfnamefont{C.~R.} \bibnamefont{Ast}}, \bibnamefont{and}
  \bibinfo{author}{\bibfnamefont{K.}~\bibnamefont{Kern}},
  \bibinfo{journal}{Nano\ Lett.} \textbf{\bibinfo{volume}{8}},
  \bibinfo{pages}{4603} (\bibinfo{year}{2008}).

\bibitem[{\citenamefont{Peres}(2010)}]{P10_3}
\bibinfo{author}{\bibfnamefont{N.~M.~R.} \bibnamefont{Peres}},
  \bibinfo{journal}{Rev.\ Mod.\ Phys.} \textbf{\bibinfo{volume}{82}},
  \bibinfo{pages}{2673} (\bibinfo{year}{2010}).

\bibitem[{\citenamefont{Ni et~al.}(2008)\citenamefont{Ni, Yu, Lu, Wang, Feng,
  and Shen}}]{NYL08}
\bibinfo{author}{\bibfnamefont{Z.~H.} \bibnamefont{Ni}},
  \bibinfo{author}{\bibfnamefont{T.}~\bibnamefont{Yu}},
  \bibinfo{author}{\bibfnamefont{Y.~H.} \bibnamefont{Lu}},
  \bibinfo{author}{\bibfnamefont{Y.~Y.} \bibnamefont{Wang}},
  \bibinfo{author}{\bibfnamefont{Y.~P.} \bibnamefont{Feng}}, \bibnamefont{and}
  \bibinfo{author}{\bibfnamefont{Z.~X.} \bibnamefont{Shen}},
  \bibinfo{journal}{ACS\ Nano} \textbf{\bibinfo{volume}{2}},
  \bibinfo{pages}{2301} (\bibinfo{year}{2008}).

\bibitem[{\citenamefont{Conrad et~al.}(2017)\citenamefont{Conrad, Wang, Nevius,
  Jinkins, Celis, Narayanan~Nair, Taleb-Ibrahimi, Tejeda, Garreau, Vlad
  et~al.}}]{CWN17}
\bibinfo{author}{\bibfnamefont{M.}~\bibnamefont{Conrad}},
  \bibinfo{author}{\bibfnamefont{F.}~\bibnamefont{Wang}},
  \bibinfo{author}{\bibfnamefont{M.}~\bibnamefont{Nevius}},
  \bibinfo{author}{\bibfnamefont{K.}~\bibnamefont{Jinkins}},
  \bibinfo{author}{\bibfnamefont{A.}~\bibnamefont{Celis}},
  \bibinfo{author}{\bibfnamefont{M.}~\bibnamefont{Narayanan~Nair}},
  \bibinfo{author}{\bibfnamefont{A.}~\bibnamefont{Taleb-Ibrahimi}},
  \bibinfo{author}{\bibfnamefont{A.}~\bibnamefont{Tejeda}},
  \bibinfo{author}{\bibfnamefont{Y.}~\bibnamefont{Garreau}},
  \bibinfo{author}{\bibfnamefont{A.}~\bibnamefont{Vlad}}, \bibnamefont{et~al.},
  \bibinfo{journal}{Nano\ Lett.} \textbf{\bibinfo{volume}{17}},
  \bibinfo{pages}{341} (\bibinfo{year}{2017}).

\bibitem[{\citenamefont{Zhou et~al.}(2007)\citenamefont{Zhou, Gweon, Fedorov,
  First, {de Heer}, Lee, Guinea, {Castro Neto}, and Lanzara}}]{ZGF07}
\bibinfo{author}{\bibfnamefont{S.~Y.} \bibnamefont{Zhou}},
  \bibinfo{author}{\bibfnamefont{G.~H.} \bibnamefont{Gweon}},
  \bibinfo{author}{\bibfnamefont{A.~V.} \bibnamefont{Fedorov}},
  \bibinfo{author}{\bibfnamefont{P.~N.} \bibnamefont{First}},
  \bibinfo{author}{\bibfnamefont{W.~A.} \bibnamefont{{de Heer}}},
  \bibinfo{author}{\bibfnamefont{D.~H.} \bibnamefont{Lee}},
  \bibinfo{author}{\bibfnamefont{F.}~\bibnamefont{Guinea}},
  \bibinfo{author}{\bibfnamefont{A.~H.} \bibnamefont{{Castro Neto}}},
  \bibnamefont{and} \bibinfo{author}{\bibfnamefont{A.}~\bibnamefont{Lanzara}},
  \bibinfo{journal}{Nat.\ Mater.} \textbf{\bibinfo{volume}{6}},
  \bibinfo{pages}{3288} (\bibinfo{year}{2007}).

\bibitem[{\citenamefont{Vita et~al.}(2014)\citenamefont{Vita, B\"ottcher, Horn,
  Voloshina, Ovcharenko, Kampen, Thissen, and Dedkov}}]{VBH14}
\bibinfo{author}{\bibfnamefont{H.}~\bibnamefont{Vita}},
  \bibinfo{author}{\bibfnamefont{S.}~\bibnamefont{B\"ottcher}},
  \bibinfo{author}{\bibfnamefont{K.}~\bibnamefont{Horn}},
  \bibinfo{author}{\bibfnamefont{E.~N.} \bibnamefont{Voloshina}},
  \bibinfo{author}{\bibfnamefont{R.~E.} \bibnamefont{Ovcharenko}},
  \bibinfo{author}{\bibfnamefont{T.}~\bibnamefont{Kampen}},
  \bibinfo{author}{\bibfnamefont{A.}~\bibnamefont{Thissen}}, \bibnamefont{and}
  \bibinfo{author}{\bibfnamefont{Y.~S.} \bibnamefont{Dedkov}},
  \bibinfo{journal}{Sci.\ Rep.} \textbf{\bibinfo{volume}{4}},
  \bibinfo{pages}{5704} (\bibinfo{year}{2014}).

\bibitem[{\citenamefont{Papagno et~al.}(2012)\citenamefont{Papagno, Rusponi,
  Sheverdyaeva, Vlaic, Etzkorn, Pacil\'e, Moras, Carbone, and Brune}}]{PRS12}
\bibinfo{author}{\bibfnamefont{M.}~\bibnamefont{Papagno}},
  \bibinfo{author}{\bibfnamefont{S.}~\bibnamefont{Rusponi}},
  \bibinfo{author}{\bibfnamefont{P.~M.} \bibnamefont{Sheverdyaeva}},
  \bibinfo{author}{\bibfnamefont{S.}~\bibnamefont{Vlaic}},
  \bibinfo{author}{\bibfnamefont{M.}~\bibnamefont{Etzkorn}},
  \bibinfo{author}{\bibfnamefont{D.}~\bibnamefont{Pacil\'e}},
  \bibinfo{author}{\bibfnamefont{P.}~\bibnamefont{Moras}},
  \bibinfo{author}{\bibfnamefont{C.}~\bibnamefont{Carbone}}, \bibnamefont{and}
  \bibinfo{author}{\bibfnamefont{H.}~\bibnamefont{Brune}},
  \bibinfo{journal}{ACS\ Nano} \textbf{\bibinfo{volume}{6}},
  \bibinfo{pages}{199} (\bibinfo{year}{2012}).

\bibitem[{\citenamefont{Varykhalov et~al.}(2012)\citenamefont{Varykhalov,
  Marchenko, S\'anchez-Barriga, Scholz, Verberck, B.~Trauzettel, Carbone, , and
  Rader}}]{VMS12}
\bibinfo{author}{\bibfnamefont{A.}~\bibnamefont{Varykhalov}},
  \bibinfo{author}{\bibfnamefont{D.}~\bibnamefont{Marchenko}},
  \bibinfo{author}{\bibfnamefont{J.}~\bibnamefont{S\'anchez-Barriga}},
  \bibinfo{author}{\bibfnamefont{M.~R.} \bibnamefont{Scholz}},
  \bibinfo{author}{\bibfnamefont{B.}~\bibnamefont{Verberck}},
  \bibinfo{author}{\bibfnamefont{T.~O.~W.} \bibnamefont{B.~Trauzettel}},
  \bibinfo{author}{\bibfnamefont{C.}~\bibnamefont{Carbone}}, ,
  \bibnamefont{and} \bibinfo{author}{\bibfnamefont{O.}~\bibnamefont{Rader}},
  \bibinfo{journal}{Phys.\ Rev.\ X} \textbf{\bibinfo{volume}{2}},
  \bibinfo{pages}{041017} (\bibinfo{year}{2012}).

\bibitem[{\citenamefont{Chen et~al.}(2015)\citenamefont{Chen, Cao, Chen,
  Pedramrazi, Haberer, de~Oteyza, Fischer, Louie, and Crommie}}]{CCC15}
\bibinfo{author}{\bibfnamefont{Y.-C.} \bibnamefont{Chen}},
  \bibinfo{author}{\bibfnamefont{T.}~\bibnamefont{Cao}},
  \bibinfo{author}{\bibfnamefont{C.}~\bibnamefont{Chen}},
  \bibinfo{author}{\bibfnamefont{Z.}~\bibnamefont{Pedramrazi}},
  \bibinfo{author}{\bibfnamefont{D.}~\bibnamefont{Haberer}},
  \bibinfo{author}{\bibfnamefont{D.~G.} \bibnamefont{de~Oteyza}},
  \bibinfo{author}{\bibfnamefont{F.~R.} \bibnamefont{Fischer}},
  \bibinfo{author}{\bibfnamefont{S.~G.} \bibnamefont{Louie}}, \bibnamefont{and}
  \bibinfo{author}{\bibfnamefont{M.~F.} \bibnamefont{Crommie}},
  \bibinfo{journal}{Nat.\ Nanotech.} \textbf{\bibinfo{volume}{10}},
  \bibinfo{pages}{156} (\bibinfo{year}{2015}).

\bibitem[{\citenamefont{{Di Giovannantonio} et~al.}(2018)\citenamefont{{Di
  Giovannantonio}, Deniz, Urgel, Widmer, Dienel, Stolz, S\'anchez-S\'anchez,
  Muntwiler, Dumslaff, Berger et~al.}}]{DDU18}
\bibinfo{author}{\bibfnamefont{M.}~\bibnamefont{{Di Giovannantonio}}},
  \bibinfo{author}{\bibfnamefont{O.}~\bibnamefont{Deniz}},
  \bibinfo{author}{\bibfnamefont{J.~I.} \bibnamefont{Urgel}},
  \bibinfo{author}{\bibfnamefont{R.}~\bibnamefont{Widmer}},
  \bibinfo{author}{\bibfnamefont{T.}~\bibnamefont{Dienel}},
  \bibinfo{author}{\bibfnamefont{S.}~\bibnamefont{Stolz}},
  \bibinfo{author}{\bibfnamefont{C.}~\bibnamefont{S\'anchez-S\'anchez}},
  \bibinfo{author}{\bibfnamefont{M.}~\bibnamefont{Muntwiler}},
  \bibinfo{author}{\bibfnamefont{T.}~\bibnamefont{Dumslaff}},
  \bibinfo{author}{\bibfnamefont{R.}~\bibnamefont{Berger}},
  \bibnamefont{et~al.}, \bibinfo{journal}{ACS\ Nano}
  \textbf{\bibinfo{volume}{12}}, \bibinfo{pages}{74} (\bibinfo{year}{2018}).

\bibitem[{\citenamefont{Bronner et~al.}(2018)\citenamefont{Bronner, Durr,
  Rizzo, Lee, Marangoni, Kalayjian, Rodriguez, Zhao, Louie, Fischer
  et~al.}}]{BDR18}
\bibinfo{author}{\bibfnamefont{C.}~\bibnamefont{Bronner}},
  \bibinfo{author}{\bibfnamefont{R.~A.} \bibnamefont{Durr}},
  \bibinfo{author}{\bibfnamefont{D.~J.} \bibnamefont{Rizzo}},
  \bibinfo{author}{\bibfnamefont{Y.-L.} \bibnamefont{Lee}},
  \bibinfo{author}{\bibfnamefont{T.}~\bibnamefont{Marangoni}},
  \bibinfo{author}{\bibfnamefont{A.~M.} \bibnamefont{Kalayjian}},
  \bibinfo{author}{\bibfnamefont{H.}~\bibnamefont{Rodriguez}},
  \bibinfo{author}{\bibfnamefont{W.}~\bibnamefont{Zhao}},
  \bibinfo{author}{\bibfnamefont{S.~G.} \bibnamefont{Louie}},
  \bibinfo{author}{\bibfnamefont{F.~R.} \bibnamefont{Fischer}},
  \bibnamefont{et~al.}, \bibinfo{journal}{ACS Nano}
  \textbf{\bibinfo{volume}{12}}, \bibinfo{pages}{2193} (\bibinfo{year}{2018}).

\bibitem[{\citenamefont{Gr\"oning et~al.}(2018)\citenamefont{Gr\"oning,
  Sh.~Wang, Pignedoli, Barin, Daniels, Cupo, Meunier, Feng, Narita, M\"ullen
  et~al.}}]{GNR_junctions}
\bibinfo{author}{\bibfnamefont{O.}~\bibnamefont{Gr\"oning}},
  \bibinfo{author}{\bibfnamefont{X.~Y.} \bibnamefont{Sh.~Wang}},
  \bibinfo{author}{\bibfnamefont{C.~A.} \bibnamefont{Pignedoli}},
  \bibinfo{author}{\bibfnamefont{G.~B.} \bibnamefont{Barin}},
  \bibinfo{author}{\bibfnamefont{C.}~\bibnamefont{Daniels}},
  \bibinfo{author}{\bibfnamefont{A.}~\bibnamefont{Cupo}},
  \bibinfo{author}{\bibfnamefont{V.}~\bibnamefont{Meunier}},
  \bibinfo{author}{\bibfnamefont{X.}~\bibnamefont{Feng}},
  \bibinfo{author}{\bibfnamefont{A.}~\bibnamefont{Narita}},
  \bibinfo{author}{\bibfnamefont{K.}~\bibnamefont{M\"ullen}},
  \bibnamefont{et~al.}, \bibinfo{journal}{Nature}
  \textbf{\bibinfo{volume}{560}}, \bibinfo{pages}{209} (\bibinfo{year}{2018}).

\bibitem[{\citenamefont{Rizzo et~al.}(2018)\citenamefont{Rizzo, Veber, Cao,
  Bronner, Chen, Zhao, Rodriguez, Louie, Crommie, and
  Fischer}}]{GNR_junctions2}
\bibinfo{author}{\bibfnamefont{D.~J.} \bibnamefont{Rizzo}},
  \bibinfo{author}{\bibfnamefont{G.}~\bibnamefont{Veber}},
  \bibinfo{author}{\bibfnamefont{T.}~\bibnamefont{Cao}},
  \bibinfo{author}{\bibfnamefont{C.}~\bibnamefont{Bronner}},
  \bibinfo{author}{\bibfnamefont{T.}~\bibnamefont{Chen}},
  \bibinfo{author}{\bibfnamefont{F.}~\bibnamefont{Zhao}},
  \bibinfo{author}{\bibfnamefont{H.}~\bibnamefont{Rodriguez}},
  \bibinfo{author}{\bibfnamefont{S.~G.} \bibnamefont{Louie}},
  \bibinfo{author}{\bibfnamefont{M.~F.} \bibnamefont{Crommie}},
  \bibnamefont{and} \bibinfo{author}{\bibfnamefont{F.~R.}
  \bibnamefont{Fischer}}, \bibinfo{journal}{Nature}
  \textbf{\bibinfo{volume}{560}}, \bibinfo{pages}{204} (\bibinfo{year}{2018}).

\bibitem[{\citenamefont{Moreno et~al.}(2018)\citenamefont{Moreno, Vilas-Varela,
  Kretz, Garcia-Lekue, Costache, Paradinas, Panighel, Ceballos, Valenzuela,
  Pe{\~n}a et~al.}}]{MVK18}
\bibinfo{author}{\bibfnamefont{C.}~\bibnamefont{Moreno}},
  \bibinfo{author}{\bibfnamefont{M.}~\bibnamefont{Vilas-Varela}},
  \bibinfo{author}{\bibfnamefont{B.}~\bibnamefont{Kretz}},
  \bibinfo{author}{\bibfnamefont{A.}~\bibnamefont{Garcia-Lekue}},
  \bibinfo{author}{\bibfnamefont{M.~V.} \bibnamefont{Costache}},
  \bibinfo{author}{\bibfnamefont{M.}~\bibnamefont{Paradinas}},
  \bibinfo{author}{\bibfnamefont{M.}~\bibnamefont{Panighel}},
  \bibinfo{author}{\bibfnamefont{G.}~\bibnamefont{Ceballos}},
  \bibinfo{author}{\bibfnamefont{S.~O.} \bibnamefont{Valenzuela}},
  \bibinfo{author}{\bibfnamefont{D.}~\bibnamefont{Pe{\~n}a}},
  \bibnamefont{et~al.}, \bibinfo{journal}{Science}
  \textbf{\bibinfo{volume}{360}}, \bibinfo{pages}{199} (\bibinfo{year}{2018}).

\bibitem[{\citenamefont{Ruffieux et~al.}(2012)\citenamefont{Ruffieux, Cai,
  Plumb, Patthey, Prezzi, Ferretti, Molinari, Feng, MÃ¼llen, Pignedoli
  et~al.}}]{RCP12}
\bibinfo{author}{\bibfnamefont{P.}~\bibnamefont{Ruffieux}},
  \bibinfo{author}{\bibfnamefont{J.}~\bibnamefont{Cai}},
  \bibinfo{author}{\bibfnamefont{N.~C.} \bibnamefont{Plumb}},
  \bibinfo{author}{\bibfnamefont{L.}~\bibnamefont{Patthey}},
  \bibinfo{author}{\bibfnamefont{D.}~\bibnamefont{Prezzi}},
  \bibinfo{author}{\bibfnamefont{A.}~\bibnamefont{Ferretti}},
  \bibinfo{author}{\bibfnamefont{E.}~\bibnamefont{Molinari}},
  \bibinfo{author}{\bibfnamefont{X.}~\bibnamefont{Feng}},
  \bibinfo{author}{\bibfnamefont{K.}~\bibnamefont{MÃ¼llen}},
  \bibinfo{author}{\bibfnamefont{C.~A.} \bibnamefont{Pignedoli}},
  \bibnamefont{et~al.}, \bibinfo{journal}{ACS\ Nano}
  \textbf{\bibinfo{volume}{6}}, \bibinfo{pages}{6930} (\bibinfo{year}{2012}).

\bibitem[{\citenamefont{Senkovskiy
  et~al.}(2018{\natexlab{a}})\citenamefont{Senkovskiy, Usachov, Fedorov,
  Haberer, Ehlen, Fischer, and Gr\"uneis}}]{SUF18}
\bibinfo{author}{\bibfnamefont{B.~V.} \bibnamefont{Senkovskiy}},
  \bibinfo{author}{\bibfnamefont{D.~Y.} \bibnamefont{Usachov}},
  \bibinfo{author}{\bibfnamefont{A.~V.} \bibnamefont{Fedorov}},
  \bibinfo{author}{\bibfnamefont{D.}~\bibnamefont{Haberer}},
  \bibinfo{author}{\bibfnamefont{N.}~\bibnamefont{Ehlen}},
  \bibinfo{author}{\bibfnamefont{F.~R.} \bibnamefont{Fischer}},
  \bibnamefont{and}
  \bibinfo{author}{\bibfnamefont{A.}~\bibnamefont{Gr\"uneis}},
  \bibinfo{journal}{2D\ Mater.} \textbf{\bibinfo{volume}{5}},
  \bibinfo{pages}{035007} (\bibinfo{year}{2018}{\natexlab{a}}).

\bibitem[{\citenamefont{Carbonell-Sanrom\'a
  et~al.}(2018)\citenamefont{Carbonell-Sanrom\'a, Garcia-Lekue, Corso, Vasseur,
  Brandimarte, Lobo-Checa, de~Oteyza, Li, Kawai, Saito et~al.}}]{CGC18}
\bibinfo{author}{\bibfnamefont{E.}~\bibnamefont{Carbonell-Sanrom\'a}},
  \bibinfo{author}{\bibfnamefont{A.}~\bibnamefont{Garcia-Lekue}},
  \bibinfo{author}{\bibfnamefont{M.}~\bibnamefont{Corso}},
  \bibinfo{author}{\bibfnamefont{G.}~\bibnamefont{Vasseur}},
  \bibinfo{author}{\bibfnamefont{P.}~\bibnamefont{Brandimarte}},
  \bibinfo{author}{\bibfnamefont{J.}~\bibnamefont{Lobo-Checa}},
  \bibinfo{author}{\bibfnamefont{D.~G.} \bibnamefont{de~Oteyza}},
  \bibinfo{author}{\bibfnamefont{J.}~\bibnamefont{Li}},
  \bibinfo{author}{\bibfnamefont{S.}~\bibnamefont{Kawai}},
  \bibinfo{author}{\bibfnamefont{S.}~\bibnamefont{Saito}},
  \bibnamefont{et~al.}, \bibinfo{journal}{J.\ Phys.\ Chem.\ C}
  \textbf{\bibinfo{volume}{122}}, \bibinfo{pages}{16092}
  (\bibinfo{year}{2018}).

\bibitem[{\citenamefont{Magda et~al.}(2014)\citenamefont{Magda, Jin,
  Hagym\'asi, Vancs\'o, Osv\'ath, Nemes-Incze, Hwang, Bir\'o, and
  Tapaszt\'o}}]{MJH14}
\bibinfo{author}{\bibfnamefont{G.~Z.} \bibnamefont{Magda}},
  \bibinfo{author}{\bibfnamefont{X.}~\bibnamefont{Jin}},
  \bibinfo{author}{\bibfnamefont{I.}~\bibnamefont{Hagym\'asi}},
  \bibinfo{author}{\bibfnamefont{P.}~\bibnamefont{Vancs\'o}},
  \bibinfo{author}{\bibfnamefont{Z.}~\bibnamefont{Osv\'ath}},
  \bibinfo{author}{\bibfnamefont{P.}~\bibnamefont{Nemes-Incze}},
  \bibinfo{author}{\bibfnamefont{C.}~\bibnamefont{Hwang}},
  \bibinfo{author}{\bibfnamefont{L.~P.} \bibnamefont{Bir\'o}},
  \bibnamefont{and}
  \bibinfo{author}{\bibfnamefont{L.}~\bibnamefont{Tapaszt\'o}},
  \bibinfo{journal}{Nature} \textbf{\bibinfo{volume}{514}},
  \bibinfo{pages}{608} (\bibinfo{year}{2014}).

\bibitem[{\citenamefont{Ihn et~al.}(2010)\citenamefont{Ihn, G\"uttinger,
  Molitor, Schnez, Schurtenberger, Jacobsen, Hellm\"uller, Frey, Dr\"oscher,
  Stampfer et~al.}}]{IGM10}
\bibinfo{author}{\bibfnamefont{T.}~\bibnamefont{Ihn}},
  \bibinfo{author}{\bibfnamefont{J.}~\bibnamefont{G\"uttinger}},
  \bibinfo{author}{\bibfnamefont{F.}~\bibnamefont{Molitor}},
  \bibinfo{author}{\bibfnamefont{S.}~\bibnamefont{Schnez}},
  \bibinfo{author}{\bibfnamefont{E.}~\bibnamefont{Schurtenberger}},
  \bibinfo{author}{\bibfnamefont{A.}~\bibnamefont{Jacobsen}},
  \bibinfo{author}{\bibfnamefont{S.}~\bibnamefont{Hellm\"uller}},
  \bibinfo{author}{\bibfnamefont{T.}~\bibnamefont{Frey}},
  \bibinfo{author}{\bibfnamefont{S.}~\bibnamefont{Dr\"oscher}},
  \bibinfo{author}{\bibfnamefont{C.}~\bibnamefont{Stampfer}},
  \bibnamefont{et~al.}, \bibinfo{journal}{Mater.\ Today}
  \textbf{\bibinfo{volume}{13}}, \bibinfo{pages}{44} (\bibinfo{year}{2010}).

\bibitem[{\citenamefont{Llinas et~al.}(2017)\citenamefont{Llinas, Fairbrother,
  {Borin Barin}, Shi, Lee, Wu, Choi, Braganza, Lear, Kau et~al.}}]{LFB17}
\bibinfo{author}{\bibfnamefont{J.~P.} \bibnamefont{Llinas}},
  \bibinfo{author}{\bibfnamefont{A.}~\bibnamefont{Fairbrother}},
  \bibinfo{author}{\bibfnamefont{G.}~\bibnamefont{{Borin Barin}}},
  \bibinfo{author}{\bibfnamefont{W.}~\bibnamefont{Shi}},
  \bibinfo{author}{\bibfnamefont{K.}~\bibnamefont{Lee}},
  \bibinfo{author}{\bibfnamefont{S.}~\bibnamefont{Wu}},
  \bibinfo{author}{\bibfnamefont{B.~Y.} \bibnamefont{Choi}},
  \bibinfo{author}{\bibfnamefont{R.}~\bibnamefont{Braganza}},
  \bibinfo{author}{\bibfnamefont{J.}~\bibnamefont{Lear}},
  \bibinfo{author}{\bibfnamefont{N.}~\bibnamefont{Kau}}, \bibnamefont{et~al.},
  \bibinfo{journal}{Nat.\ Commun.} \textbf{\bibinfo{volume}{8}},
  \bibinfo{pages}{633} (\bibinfo{year}{2017}).

\bibitem[{\citenamefont{Celis et~al.}(2016)\citenamefont{Celis, Nair,
  Taleb-Ibrahimi, Conrad, Berger, de~Heer, and Tejeda}}]{CNT16}
\bibinfo{author}{\bibfnamefont{A.}~\bibnamefont{Celis}},
  \bibinfo{author}{\bibfnamefont{M.~N.} \bibnamefont{Nair}},
  \bibinfo{author}{\bibfnamefont{A.}~\bibnamefont{Taleb-Ibrahimi}},
  \bibinfo{author}{\bibfnamefont{E.~H.} \bibnamefont{Conrad}},
  \bibinfo{author}{\bibfnamefont{C.}~\bibnamefont{Berger}},
  \bibinfo{author}{\bibfnamefont{W.~A.} \bibnamefont{de~Heer}},
  \bibnamefont{and} \bibinfo{author}{\bibfnamefont{A.}~\bibnamefont{Tejeda}},
  \bibinfo{journal}{J.\ Phys.\ D:\ Appl.\ Phys.} \textbf{\bibinfo{volume}{49}},
  \bibinfo{pages}{143001} (\bibinfo{year}{2016}).

\bibitem[{\citenamefont{Wallace}(1947)}]{W1947}
\bibinfo{author}{\bibfnamefont{P.~R.} \bibnamefont{Wallace}},
  \bibinfo{journal}{Phys.\ Rev.} \textbf{\bibinfo{volume}{71}},
  \bibinfo{pages}{622} (\bibinfo{year}{1947}).

\bibitem[{\citenamefont{Son et~al.}(2006{\natexlab{a}})\citenamefont{Son,
  Cohen, and Louie}}]{SCL06_2}
\bibinfo{author}{\bibfnamefont{Y.-W.} \bibnamefont{Son}},
  \bibinfo{author}{\bibfnamefont{M.~L.} \bibnamefont{Cohen}}, \bibnamefont{and}
  \bibinfo{author}{\bibfnamefont{S.~G.} \bibnamefont{Louie}},
  \bibinfo{journal}{Nature} \textbf{\bibinfo{volume}{444}},
  \bibinfo{pages}{348} (\bibinfo{year}{2006}{\natexlab{a}}).

\bibitem[{\citenamefont{Kimouche et~al.}(2015)\citenamefont{Kimouche, Ervasti,
  Drost, Halonen, Harju, Joensuu, Sainio, and Liljeroth}}]{KED15}
\bibinfo{author}{\bibfnamefont{A.}~\bibnamefont{Kimouche}},
  \bibinfo{author}{\bibfnamefont{M.~M.} \bibnamefont{Ervasti}},
  \bibinfo{author}{\bibfnamefont{R.}~\bibnamefont{Drost}},
  \bibinfo{author}{\bibfnamefont{S.}~\bibnamefont{Halonen}},
  \bibinfo{author}{\bibfnamefont{A.}~\bibnamefont{Harju}},
  \bibinfo{author}{\bibfnamefont{P.~M.} \bibnamefont{Joensuu}},
  \bibinfo{author}{\bibfnamefont{J.}~\bibnamefont{Sainio}}, \bibnamefont{and}
  \bibinfo{author}{\bibfnamefont{P.}~\bibnamefont{Liljeroth}},
  \bibinfo{journal}{Nat.\ Commun.} \textbf{\bibinfo{volume}{6}},
  \bibinfo{pages}{10177} (\bibinfo{year}{2015}).

\bibitem[{\citenamefont{Merino-D\'{\i}ez
  et~al.}(2017)\citenamefont{Merino-D\'{\i}ez, Garcia-Lekue,
  Carbonell-Sanrom\'a, Li, Corso, Colazzo, Sedona, S\'anchez-Portal, Pascual,
  and de~Oteyza}}]{MGC17}
\bibinfo{author}{\bibfnamefont{N.}~\bibnamefont{Merino-D\'{\i}ez}},
  \bibinfo{author}{\bibfnamefont{A.}~\bibnamefont{Garcia-Lekue}},
  \bibinfo{author}{\bibfnamefont{E.}~\bibnamefont{Carbonell-Sanrom\'a}},
  \bibinfo{author}{\bibfnamefont{J.}~\bibnamefont{Li}},
  \bibinfo{author}{\bibfnamefont{M.}~\bibnamefont{Corso}},
  \bibinfo{author}{\bibfnamefont{L.}~\bibnamefont{Colazzo}},
  \bibinfo{author}{\bibfnamefont{F.}~\bibnamefont{Sedona}},
  \bibinfo{author}{\bibfnamefont{D.}~\bibnamefont{S\'anchez-Portal}},
  \bibinfo{author}{\bibfnamefont{J.~I.} \bibnamefont{Pascual}},
  \bibnamefont{and} \bibinfo{author}{\bibfnamefont{D.~G.}
  \bibnamefont{de~Oteyza}}, \bibinfo{journal}{ACS\ Nano}
  \textbf{\bibinfo{volume}{11}}, \bibinfo{pages}{11661} (\bibinfo{year}{2017}).

\bibitem[{\citenamefont{S\"ode et~al.}(2015)\citenamefont{S\"ode, Talirz,
  Gr\"oning, Pignedoli, Berger, Feng, M\"ullen, Fasel, and Ruffieux}}]{STG15}
\bibinfo{author}{\bibfnamefont{H.}~\bibnamefont{S\"ode}},
  \bibinfo{author}{\bibfnamefont{L.}~\bibnamefont{Talirz}},
  \bibinfo{author}{\bibfnamefont{O.}~\bibnamefont{Gr\"oning}},
  \bibinfo{author}{\bibfnamefont{C.~A.} \bibnamefont{Pignedoli}},
  \bibinfo{author}{\bibfnamefont{R.}~\bibnamefont{Berger}},
  \bibinfo{author}{\bibfnamefont{X.}~\bibnamefont{Feng}},
  \bibinfo{author}{\bibfnamefont{K.}~\bibnamefont{M\"ullen}},
  \bibinfo{author}{\bibfnamefont{R.}~\bibnamefont{Fasel}}, \bibnamefont{and}
  \bibinfo{author}{\bibfnamefont{P.}~\bibnamefont{Ruffieux}},
  \bibinfo{journal}{Phys.\ Rev.\ B} \textbf{\bibinfo{volume}{91}},
  \bibinfo{pages}{045429} (\bibinfo{year}{2015}).

\bibitem[{\citenamefont{Li et~al.}(2014)\citenamefont{Li, Chen, Weinert, and
  Li}}]{LCW14}
\bibinfo{author}{\bibfnamefont{Y.}~\bibnamefont{Li}},
  \bibinfo{author}{\bibfnamefont{M.}~\bibnamefont{Chen}},
  \bibinfo{author}{\bibfnamefont{M.}~\bibnamefont{Weinert}}, \bibnamefont{and}
  \bibinfo{author}{\bibfnamefont{L.}~\bibnamefont{Li}}, \bibinfo{journal}{Nat.\
  Commun.} \textbf{\bibinfo{volume}{5}}, \bibinfo{pages}{4311}
  (\bibinfo{year}{2014}).

\bibitem[{\citenamefont{Kittel}(1987)}]{K1987}
\bibinfo{author}{\bibfnamefont{C.}~\bibnamefont{Kittel}},
  \emph{\bibinfo{title}{Quantum theory of solids}} (\bibinfo{publisher}{Wiley},
  \bibinfo{address}{New York}, \bibinfo{year}{1987}).

\bibitem[{\citenamefont{Ashcroft and Mermin}(1976)}]{AM1976}
\bibinfo{author}{\bibfnamefont{N.~W.} \bibnamefont{Ashcroft}} \bibnamefont{and}
  \bibinfo{author}{\bibfnamefont{N.~D.} \bibnamefont{Mermin}},
  \emph{\bibinfo{title}{Solid State Physics}} (\bibinfo{publisher}{Harcourt
  College Publishers}, \bibinfo{address}{Philadelphia}, \bibinfo{year}{1976}).

\bibitem[{\citenamefont{Mugarza et~al.}(2001)\citenamefont{Mugarza, Mascaraque,
  {V. P\'{e}rez-Dieste}, Repain, Rousset, {Garc\'{\i}a de Abajo}, and
  Ortega}}]{paper030}
\bibinfo{author}{\bibfnamefont{A.}~\bibnamefont{Mugarza}},
  \bibinfo{author}{\bibfnamefont{A.}~\bibnamefont{Mascaraque}},
  \bibinfo{author}{\bibnamefont{{V. P\'{e}rez-Dieste}}},
  \bibinfo{author}{\bibfnamefont{V.}~\bibnamefont{Repain}},
  \bibinfo{author}{\bibfnamefont{S.}~\bibnamefont{Rousset}},
  \bibinfo{author}{\bibfnamefont{F.~J.} \bibnamefont{{Garc\'{\i}a de Abajo}}},
  \bibnamefont{and} \bibinfo{author}{\bibfnamefont{J.~E.}
  \bibnamefont{Ortega}}, \bibinfo{journal}{Phys.\ Rev.\ Lett.}
  \textbf{\bibinfo{volume}{87}}, \bibinfo{pages}{107601}
  (\bibinfo{year}{2001}).

\bibitem[{\citenamefont{Ortega and {Garc\'{\i}a de Abajo}}(2007)}]{comm001}
\bibinfo{author}{\bibfnamefont{J.~E.} \bibnamefont{Ortega}} \bibnamefont{and}
  \bibinfo{author}{\bibfnamefont{F.~J.} \bibnamefont{{Garc\'{\i}a de Abajo}}},
  \bibinfo{journal}{Nat.\ Nanotech.} \textbf{\bibinfo{volume}{2}},
  \bibinfo{pages}{601} (\bibinfo{year}{2007}).

\bibitem[{\citenamefont{Piquero-Zulaica
  et~al.}(2017)\citenamefont{Piquero-Zulaica, Lobo-Checa, Sadeghi, {Abd
  El-Fattah}, Mitsui, Okamoto, Pawlak, Meier, Arnau, Ortega et~al.}}]{PLS17}
\bibinfo{author}{\bibfnamefont{I.}~\bibnamefont{Piquero-Zulaica}},
  \bibinfo{author}{\bibfnamefont{J.}~\bibnamefont{Lobo-Checa}},
  \bibinfo{author}{\bibfnamefont{A.}~\bibnamefont{Sadeghi}},
  \bibinfo{author}{\bibfnamefont{Z.~M.} \bibnamefont{{Abd El-Fattah}}},
  \bibinfo{author}{\bibfnamefont{C.}~\bibnamefont{Mitsui}},
  \bibinfo{author}{\bibfnamefont{T.}~\bibnamefont{Okamoto}},
  \bibinfo{author}{\bibfnamefont{R.}~\bibnamefont{Pawlak}},
  \bibinfo{author}{\bibfnamefont{T.}~\bibnamefont{Meier}},
  \bibinfo{author}{\bibfnamefont{A.}~\bibnamefont{Arnau}},
  \bibinfo{author}{\bibfnamefont{J.~E.} \bibnamefont{Ortega}},
  \bibnamefont{et~al.}, \bibinfo{journal}{Nat.\ Commun.}
  \textbf{\bibinfo{volume}{8}}, \bibinfo{pages}{787} (\bibinfo{year}{2017}).

\bibitem[{\citenamefont{Malterre et~al.}(2011)\citenamefont{Malterre, Kierren,
  Fagot-Revurat, Didiot, {Garc\'{\i}a de Abajo}, Schiller, Cord\'on, and
  Ortega}}]{paper170}
\bibinfo{author}{\bibfnamefont{D.}~\bibnamefont{Malterre}},
  \bibinfo{author}{\bibfnamefont{B.}~\bibnamefont{Kierren}},
  \bibinfo{author}{\bibfnamefont{Y.}~\bibnamefont{Fagot-Revurat}},
  \bibinfo{author}{\bibfnamefont{C.}~\bibnamefont{Didiot}},
  \bibinfo{author}{\bibfnamefont{F.~J.} \bibnamefont{{Garc\'{\i}a de Abajo}}},
  \bibinfo{author}{\bibfnamefont{F.}~\bibnamefont{Schiller}},
  \bibinfo{author}{\bibfnamefont{J.}~\bibnamefont{Cord\'on}}, \bibnamefont{and}
  \bibinfo{author}{\bibfnamefont{J.~E.} \bibnamefont{Ortega}},
  \bibinfo{journal}{New\ J.\ Phys.} \textbf{\bibinfo{volume}{13}},
  \bibinfo{pages}{013026} (\bibinfo{year}{2011}).

\bibitem[{\citenamefont{{Abd El-Fattah} et~al.}(2011)\citenamefont{{Abd
  El-Fattah}, Matena, Corso, {Garc\'{\i}a de Abajo}, Schiller, and
  Ortega}}]{paper175}
\bibinfo{author}{\bibfnamefont{Z.~M.} \bibnamefont{{Abd El-Fattah}}},
  \bibinfo{author}{\bibfnamefont{M.}~\bibnamefont{Matena}},
  \bibinfo{author}{\bibfnamefont{M.}~\bibnamefont{Corso}},
  \bibinfo{author}{\bibfnamefont{F.~J.} \bibnamefont{{Garc\'{\i}a de Abajo}}},
  \bibinfo{author}{\bibfnamefont{F.}~\bibnamefont{Schiller}}, \bibnamefont{and}
  \bibinfo{author}{\bibfnamefont{J.~E.} \bibnamefont{Ortega}},
  \bibinfo{journal}{Phys.\ Rev.\ Lett.} \textbf{\bibinfo{volume}{107}},
  \bibinfo{pages}{066803} (\bibinfo{year}{2011}).

\bibitem[{\citenamefont{Gomes et~al.}(2012)\citenamefont{Gomes, Mar, Ko,
  Guinea, and Manoharan}}]{GMK12}
\bibinfo{author}{\bibfnamefont{K.~K.} \bibnamefont{Gomes}},
  \bibinfo{author}{\bibfnamefont{W.}~\bibnamefont{Mar}},
  \bibinfo{author}{\bibfnamefont{W.}~\bibnamefont{Ko}},
  \bibinfo{author}{\bibfnamefont{F.}~\bibnamefont{Guinea}}, \bibnamefont{and}
  \bibinfo{author}{\bibfnamefont{H.~C.} \bibnamefont{Manoharan}},
  \bibinfo{journal}{Nature} \textbf{\bibinfo{volume}{483}},
  \bibinfo{pages}{306} (\bibinfo{year}{2012}).

\bibitem[{car(2000)}]{carbon_2s}
\emph{\bibinfo{title}{NIST X-ray Photoelectron Spectroscopy Database}}, NIST
  Standard Reference Database Number 20 (\bibinfo{publisher}{National Institute
  of Standards and Technology}, \bibinfo{year}{2000}).

\bibitem[{\citenamefont{Puschnig and L\"uftner}(2015)}]{PL15}
\bibinfo{author}{\bibfnamefont{P.}~\bibnamefont{Puschnig}} \bibnamefont{and}
  \bibinfo{author}{\bibfnamefont{D.}~\bibnamefont{L\"uftner}},
  \bibinfo{journal}{J.\ Electr.\ Spectrosc.\ Relat.\ Phenom.}
  \textbf{\bibinfo{volume}{200}}, \bibinfo{pages}{193} (\bibinfo{year}{2015}).

\bibitem[{\citenamefont{Rybkina et~al.}(2013)\citenamefont{Rybkina, Rybkin,
  Fedorov, Usachov, Yachmenev, Marchenko, Vilkov, Nelyubov, Adamchuk, and
  Shikin}}]{RRF13}
\bibinfo{author}{\bibfnamefont{A.~A.} \bibnamefont{Rybkina}},
  \bibinfo{author}{\bibfnamefont{A.~G.} \bibnamefont{Rybkin}},
  \bibinfo{author}{\bibfnamefont{A.~V.} \bibnamefont{Fedorov}},
  \bibinfo{author}{\bibfnamefont{D.~Y.} \bibnamefont{Usachov}},
  \bibinfo{author}{\bibfnamefont{M.~E.} \bibnamefont{Yachmenev}},
  \bibinfo{author}{\bibfnamefont{D.~E.} \bibnamefont{Marchenko}},
  \bibinfo{author}{\bibfnamefont{O.~Y.} \bibnamefont{Vilkov}},
  \bibinfo{author}{\bibfnamefont{A.~V.} \bibnamefont{Nelyubov}},
  \bibinfo{author}{\bibfnamefont{V.~K.} \bibnamefont{Adamchuk}},
  \bibnamefont{and} \bibinfo{author}{\bibfnamefont{A.~M.}
  \bibnamefont{Shikin}}, \bibinfo{journal}{Surf.\ Sci.}
  \textbf{\bibinfo{volume}{609}}, \bibinfo{pages}{7} (\bibinfo{year}{2013}).

\bibitem[{\citenamefont{Dedkov and Voloshina}(2015)}]{DV15}
\bibinfo{author}{\bibfnamefont{Y.}~\bibnamefont{Dedkov}} \bibnamefont{and}
  \bibinfo{author}{\bibfnamefont{E.}~\bibnamefont{Voloshina}},
  \bibinfo{journal}{J.\ Phys.:\ Condens.\ Matter}
  \textbf{\bibinfo{volume}{27}}, \bibinfo{pages}{303002}
  (\bibinfo{year}{2015}).

\bibitem[{\citenamefont{Kretinin et~al.}(2013)\citenamefont{Kretinin, Yu,
  Jalil, Cao, Withers, Mishchenko, Katsnelson, Novoselov, Geim, and
  Guinea}}]{KYJ13}
\bibinfo{author}{\bibfnamefont{A.}~\bibnamefont{Kretinin}},
  \bibinfo{author}{\bibfnamefont{G.~L.} \bibnamefont{Yu}},
  \bibinfo{author}{\bibfnamefont{R.}~\bibnamefont{Jalil}},
  \bibinfo{author}{\bibfnamefont{Y.}~\bibnamefont{Cao}},
  \bibinfo{author}{\bibfnamefont{F.}~\bibnamefont{Withers}},
  \bibinfo{author}{\bibfnamefont{A.}~\bibnamefont{Mishchenko}},
  \bibinfo{author}{\bibfnamefont{M.~I.} \bibnamefont{Katsnelson}},
  \bibinfo{author}{\bibfnamefont{K.~S.} \bibnamefont{Novoselov}},
  \bibinfo{author}{\bibfnamefont{A.~K.} \bibnamefont{Geim}}, \bibnamefont{and}
  \bibinfo{author}{\bibfnamefont{F.}~\bibnamefont{Guinea}},
  \bibinfo{journal}{Phys.\ Rev.\ B} \textbf{\bibinfo{volume}{88}},
  \bibinfo{pages}{165427} (\bibinfo{year}{2013}).

\bibitem[{\citenamefont{Reich et~al.}(2002)\citenamefont{Reich, Maultzsch,
  Thomsen, and Ordej\'on}}]{RMT02}
\bibinfo{author}{\bibfnamefont{S.}~\bibnamefont{Reich}},
  \bibinfo{author}{\bibfnamefont{J.}~\bibnamefont{Maultzsch}},
  \bibinfo{author}{\bibfnamefont{C.}~\bibnamefont{Thomsen}}, \bibnamefont{and}
  \bibinfo{author}{\bibfnamefont{P.}~\bibnamefont{Ordej\'on}},
  \bibinfo{journal}{Phys.\ Rev.\ B} \textbf{\bibinfo{volume}{66}},
  \bibinfo{pages}{035412} (\bibinfo{year}{2002}).

\bibitem[{\citenamefont{Fedorov et~al.}(2014)\citenamefont{Fedorov, Verbitskiy,
  Haberer, Struzzi, Petaccia, Usachov, Vilkov, Vyalikh, Fink, Knupfer
  et~al.}}]{FVH14}
\bibinfo{author}{\bibfnamefont{A.~V.} \bibnamefont{Fedorov}},
  \bibinfo{author}{\bibfnamefont{N.~I.} \bibnamefont{Verbitskiy}},
  \bibinfo{author}{\bibfnamefont{D.}~\bibnamefont{Haberer}},
  \bibinfo{author}{\bibfnamefont{C.}~\bibnamefont{Struzzi}},
  \bibinfo{author}{\bibfnamefont{L.}~\bibnamefont{Petaccia}},
  \bibinfo{author}{\bibfnamefont{D.}~\bibnamefont{Usachov}},
  \bibinfo{author}{\bibfnamefont{O.~Y.} \bibnamefont{Vilkov}},
  \bibinfo{author}{\bibfnamefont{D.~V.} \bibnamefont{Vyalikh}},
  \bibinfo{author}{\bibfnamefont{J.}~\bibnamefont{Fink}},
  \bibinfo{author}{\bibfnamefont{M.}~\bibnamefont{Knupfer}},
  \bibnamefont{et~al.}, \bibinfo{journal}{Nat.\ Commun.}
  \textbf{\bibinfo{volume}{5}}, \bibinfo{pages}{3257} (\bibinfo{year}{2014}).

\bibitem[{\citenamefont{Ulstrup et~al.}(2018)\citenamefont{Ulstrup, Lacovig,
  Orlando, Lizzit, Bignardi, Dalmiglio, Bianchi, Mazzola, Baraldi, Larciprete
  et~al.}}]{ULO18}
\bibinfo{author}{\bibfnamefont{S.}~\bibnamefont{Ulstrup}},
  \bibinfo{author}{\bibfnamefont{P.}~\bibnamefont{Lacovig}},
  \bibinfo{author}{\bibfnamefont{F.}~\bibnamefont{Orlando}},
  \bibinfo{author}{\bibfnamefont{D.}~\bibnamefont{Lizzit}},
  \bibinfo{author}{\bibfnamefont{L.}~\bibnamefont{Bignardi}},
  \bibinfo{author}{\bibfnamefont{M.}~\bibnamefont{Dalmiglio}},
  \bibinfo{author}{\bibfnamefont{M.}~\bibnamefont{Bianchi}},
  \bibinfo{author}{\bibfnamefont{F.}~\bibnamefont{Mazzola}},
  \bibinfo{author}{\bibfnamefont{A.}~\bibnamefont{Baraldi}},
  \bibinfo{author}{\bibfnamefont{R.}~\bibnamefont{Larciprete}},
  \bibnamefont{et~al.}, \bibinfo{journal}{Surf.\ Sci.}  (\bibinfo{year}{2018}),
  \urlprefix\url{https://doi.org/10.1016/j.susc.2018.03.017}.

\bibitem[{\citenamefont{Varykhalov et~al.}(2008)\citenamefont{Varykhalov,
  S\'anchez-Barriga, Shikin, Biswas, Vescovo, Rybkin, Marchenko, and
  Rader}}]{VSS08}
\bibinfo{author}{\bibfnamefont{A.}~\bibnamefont{Varykhalov}},
  \bibinfo{author}{\bibfnamefont{J.}~\bibnamefont{S\'anchez-Barriga}},
  \bibinfo{author}{\bibfnamefont{A.~M.} \bibnamefont{Shikin}},
  \bibinfo{author}{\bibfnamefont{C.}~\bibnamefont{Biswas}},
  \bibinfo{author}{\bibfnamefont{E.}~\bibnamefont{Vescovo}},
  \bibinfo{author}{\bibfnamefont{A.}~\bibnamefont{Rybkin}},
  \bibinfo{author}{\bibfnamefont{D.}~\bibnamefont{Marchenko}},
  \bibnamefont{and} \bibinfo{author}{\bibfnamefont{O.}~\bibnamefont{Rader}},
  \bibinfo{journal}{Phys.\ Rev.\ Lett.} \textbf{\bibinfo{volume}{101}},
  \bibinfo{pages}{157601} (\bibinfo{year}{2008}).

\bibitem[{\citenamefont{Kralj et~al.}(2011)\citenamefont{Kralj, Pletikosi\'{c},
  Petrovi\'{c}, Pervan, Milun, N'Diaye, Busse, Michely, Fujii, and
  Vobornik}}]{KPP11}
\bibinfo{author}{\bibfnamefont{M.}~\bibnamefont{Kralj}},
  \bibinfo{author}{\bibfnamefont{I.}~\bibnamefont{Pletikosi\'{c}}},
  \bibinfo{author}{\bibfnamefont{M.}~\bibnamefont{Petrovi\'{c}}},
  \bibinfo{author}{\bibfnamefont{P.}~\bibnamefont{Pervan}},
  \bibinfo{author}{\bibfnamefont{M.}~\bibnamefont{Milun}},
  \bibinfo{author}{\bibfnamefont{A.~T.} \bibnamefont{N'Diaye}},
  \bibinfo{author}{\bibfnamefont{C.}~\bibnamefont{Busse}},
  \bibinfo{author}{\bibfnamefont{T.}~\bibnamefont{Michely}},
  \bibinfo{author}{\bibfnamefont{J.}~\bibnamefont{Fujii}}, \bibnamefont{and}
  \bibinfo{author}{\bibfnamefont{I.}~\bibnamefont{Vobornik}},
  \bibinfo{journal}{Phys.\ Rev.\ B} \textbf{\bibinfo{volume}{84}},
  \bibinfo{pages}{075427} (\bibinfo{year}{2011}).

\bibitem[{\citenamefont{Brugger et~al.}(2009)\citenamefont{Brugger, G\"unther,
  Wang, Dil, Bocquet, Osterwalder, Wintterlin, and Greber}}]{BGW09}
\bibinfo{author}{\bibfnamefont{T.}~\bibnamefont{Brugger}},
  \bibinfo{author}{\bibfnamefont{S.}~\bibnamefont{G\"unther}},
  \bibinfo{author}{\bibfnamefont{B.}~\bibnamefont{Wang}},
  \bibinfo{author}{\bibfnamefont{J.~H.} \bibnamefont{Dil}},
  \bibinfo{author}{\bibfnamefont{M.-L.} \bibnamefont{Bocquet}},
  \bibinfo{author}{\bibfnamefont{J.}~\bibnamefont{Osterwalder}},
  \bibinfo{author}{\bibfnamefont{J.}~\bibnamefont{Wintterlin}},
  \bibnamefont{and} \bibinfo{author}{\bibfnamefont{T.}~\bibnamefont{Greber}},
  \bibinfo{journal}{Phys.\ Rev.\ B} \textbf{\bibinfo{volume}{79}},
  \bibinfo{pages}{045407} (\bibinfo{year}{2009}).

\bibitem[{\citenamefont{Balog et~al.}(2010)\citenamefont{Balog, J\o{}rgensen,
  Nilsson, Andersen, Rienks, Bianchi, Fanetti, Lagsgaard, Baraldi, Lizzit
  et~al.}}]{BJN10}
\bibinfo{author}{\bibfnamefont{R.}~\bibnamefont{Balog}},
  \bibinfo{author}{\bibfnamefont{B.}~\bibnamefont{J\o{}rgensen}},
  \bibinfo{author}{\bibfnamefont{L.}~\bibnamefont{Nilsson}},
  \bibinfo{author}{\bibfnamefont{M.}~\bibnamefont{Andersen}},
  \bibinfo{author}{\bibfnamefont{E.}~\bibnamefont{Rienks}},
  \bibinfo{author}{\bibfnamefont{M.}~\bibnamefont{Bianchi}},
  \bibinfo{author}{\bibfnamefont{M.}~\bibnamefont{Fanetti}},
  \bibinfo{author}{\bibfnamefont{E.}~\bibnamefont{Lagsgaard}},
  \bibinfo{author}{\bibfnamefont{A.}~\bibnamefont{Baraldi}},
  \bibinfo{author}{\bibfnamefont{S.}~\bibnamefont{Lizzit}},
  \bibnamefont{et~al.}, \bibinfo{journal}{Nat.\ Mater.}
  \textbf{\bibinfo{volume}{9}}, \bibinfo{pages}{315} (\bibinfo{year}{2010}).

\bibitem[{\citenamefont{Vasseur et~al.}(2016)\citenamefont{Vasseur,
  Fagot-Revurat, Sicot, Kierren, Moreau, Malterre, Cardenas, Galeotti,
  Lipton-Duffin, Rosei et~al.}}]{VFS16}
\bibinfo{author}{\bibfnamefont{G.}~\bibnamefont{Vasseur}},
  \bibinfo{author}{\bibfnamefont{Y.}~\bibnamefont{Fagot-Revurat}},
  \bibinfo{author}{\bibfnamefont{M.}~\bibnamefont{Sicot}},
  \bibinfo{author}{\bibfnamefont{B.}~\bibnamefont{Kierren}},
  \bibinfo{author}{\bibfnamefont{L.}~\bibnamefont{Moreau}},
  \bibinfo{author}{\bibfnamefont{D.}~\bibnamefont{Malterre}},
  \bibinfo{author}{\bibfnamefont{L.}~\bibnamefont{Cardenas}},
  \bibinfo{author}{\bibfnamefont{G.}~\bibnamefont{Galeotti}},
  \bibinfo{author}{\bibfnamefont{J.}~\bibnamefont{Lipton-Duffin}},
  \bibinfo{author}{\bibfnamefont{F.}~\bibnamefont{Rosei}},
  \bibnamefont{et~al.}, \bibinfo{journal}{Nat.\ Commun.}
  \textbf{\bibinfo{volume}{7}}, \bibinfo{pages}{10235} (\bibinfo{year}{2016}).

\bibitem[{foo()}]{footnote}
\bibinfo{note}{Vicinal surfaces are frequently used for ARPES measurement,
  because they help to align graphene nanoribbons and define the $k_{||}$
  direction. However, the nanoribbon plane could be tilted a few degrees with
  respect to the vicinal surface plane, leading to an uncertain normal emission
  geometry, and hence to a sizeable $k_{\perp}$ projection. For example, using
  a $\alpha=5^{\circ}$ vicinal surface and He I excitation energy ($h\nu$=21.2
  eV), $k_{\perp}$ could vary by as much as $0.51\times \sqrt{h\nu-\phi}\times
  \sin \alpha \sim$0.2 \AA$^{-1}$, where ($\phi$ stands for the work function),
  leading to a confusing assignment of the detected nanoribbon bands.}

\bibitem[{\citenamefont{Piquero-Zulaica
  et~al.}(unpublished)\citenamefont{Piquero-Zulaica, Garcia-Lekue, Colazzo,
  Krug, Sabri, {Abd El-Fattah}, Gottfried, de~Oteyza, Ortega, and
  Lobo-Checa}}]{DFT_orbitals}
\bibinfo{author}{\bibfnamefont{I.}~\bibnamefont{Piquero-Zulaica}},
  \bibinfo{author}{\bibfnamefont{A.}~\bibnamefont{Garcia-Lekue}},
  \bibinfo{author}{\bibfnamefont{L.}~\bibnamefont{Colazzo}},
  \bibinfo{author}{\bibfnamefont{C.~K.} \bibnamefont{Krug}},
  \bibinfo{author}{\bibfnamefont{M.}~\bibnamefont{Sabri}},
  \bibinfo{author}{\bibfnamefont{Z.~M.} \bibnamefont{{Abd El-Fattah}}},
  \bibinfo{author}{\bibfnamefont{J.~M.} \bibnamefont{Gottfried}},
  \bibinfo{author}{\bibfnamefont{D.~G.} \bibnamefont{de~Oteyza}},
  \bibinfo{author}{\bibfnamefont{J.~E.} \bibnamefont{Ortega}},
  \bibnamefont{and}
  \bibinfo{author}{\bibfnamefont{J.}~\bibnamefont{Lobo-Checa}}
  (\bibinfo{year}{unpublished}).

\bibitem[{\citenamefont{Ruffieux et~al.}(2016)\citenamefont{Ruffieux, Wang,
  Yang, S\'anchez-S\'anchez, Liu, Dienel, Talirz, Shinde, Pignedoli, Passerone
  et~al.}}]{FWY16}
\bibinfo{author}{\bibfnamefont{P.}~\bibnamefont{Ruffieux}},
  \bibinfo{author}{\bibfnamefont{S.}~\bibnamefont{Wang}},
  \bibinfo{author}{\bibfnamefont{B.}~\bibnamefont{Yang}},
  \bibinfo{author}{\bibfnamefont{C.}~\bibnamefont{S\'anchez-S\'anchez}},
  \bibinfo{author}{\bibfnamefont{J.}~\bibnamefont{Liu}},
  \bibinfo{author}{\bibfnamefont{T.}~\bibnamefont{Dienel}},
  \bibinfo{author}{\bibfnamefont{L.}~\bibnamefont{Talirz}},
  \bibinfo{author}{\bibfnamefont{P.}~\bibnamefont{Shinde}},
  \bibinfo{author}{\bibfnamefont{C.~A.} \bibnamefont{Pignedoli}},
  \bibinfo{author}{\bibfnamefont{D.}~\bibnamefont{Passerone}},
  \bibnamefont{et~al.}, \bibinfo{journal}{Nature}
  \textbf{\bibinfo{volume}{531}}, \bibinfo{pages}{489} (\bibinfo{year}{2016}).

\bibitem[{\citenamefont{Son et~al.}(2006{\natexlab{b}})\citenamefont{Son,
  Cohen, and Louie}}]{SCL06}
\bibinfo{author}{\bibfnamefont{Y.~W.} \bibnamefont{Son}},
  \bibinfo{author}{\bibfnamefont{M.~L.} \bibnamefont{Cohen}}, \bibnamefont{and}
  \bibinfo{author}{\bibfnamefont{S.~G.} \bibnamefont{Louie}},
  \bibinfo{journal}{Phys.\ Rev.\ Lett.} \textbf{\bibinfo{volume}{97}},
  \bibinfo{pages}{216803} (\bibinfo{year}{2006}{\natexlab{b}}).

\bibitem[{\citenamefont{Yang et~al.}(2008)\citenamefont{Yang, Cohen, and
  Louie}}]{YCL08}
\bibinfo{author}{\bibfnamefont{L.}~\bibnamefont{Yang}},
  \bibinfo{author}{\bibfnamefont{M.~L.} \bibnamefont{Cohen}}, \bibnamefont{and}
  \bibinfo{author}{\bibfnamefont{S.~G.} \bibnamefont{Louie}},
  \bibinfo{journal}{Phys.\ Rev.\ Lett.} \textbf{\bibinfo{volume}{101}},
  \bibinfo{pages}{186401} (\bibinfo{year}{2008}).

\bibitem[{\citenamefont{{Garc\'{\i}a de Abajo}
  et~al.}(2010)\citenamefont{{Garc\'{\i}a de Abajo}, Cord\'on, Corso, Schiller,
  and Ortega}}]{paper153}
\bibinfo{author}{\bibfnamefont{F.~J.} \bibnamefont{{Garc\'{\i}a de Abajo}}},
  \bibinfo{author}{\bibfnamefont{J.}~\bibnamefont{Cord\'on}},
  \bibinfo{author}{\bibfnamefont{M.}~\bibnamefont{Corso}},
  \bibinfo{author}{\bibfnamefont{F.}~\bibnamefont{Schiller}}, \bibnamefont{and}
  \bibinfo{author}{\bibfnamefont{J.~E.} \bibnamefont{Ortega}},
  \bibinfo{journal}{Nanoscale} \textbf{\bibinfo{volume}{2}},
  \bibinfo{pages}{717} (\bibinfo{year}{2010}).

\bibitem[{\citenamefont{{Abd El-Fattah} et~al.}(2017)\citenamefont{{Abd
  El-Fattah}, Kher-Elden, Yassin, El-Okr, Ortega, and {Garc\'{\i}a de
  Abajo}}}]{paper301}
\bibinfo{author}{\bibfnamefont{Z.~M.} \bibnamefont{{Abd El-Fattah}}},
  \bibinfo{author}{\bibfnamefont{M.~A.} \bibnamefont{Kher-Elden}},
  \bibinfo{author}{\bibfnamefont{O.}~\bibnamefont{Yassin}},
  \bibinfo{author}{\bibfnamefont{M.~M.} \bibnamefont{El-Okr}},
  \bibinfo{author}{\bibfnamefont{J.~E.} \bibnamefont{Ortega}},
  \bibnamefont{and} \bibinfo{author}{\bibfnamefont{F.~J.}
  \bibnamefont{{Garc\'{\i}a de Abajo}}}, \bibinfo{journal}{J.\ Appl.\ Phys.}
  \textbf{\bibinfo{volume}{122}}, \bibinfo{pages}{195306}
  (\bibinfo{year}{2017}).

\bibitem[{\citenamefont{Leykam et~al.}(2018)\citenamefont{Leykam, Andreanov,
  and Flach}}]{LAF18}
\bibinfo{author}{\bibfnamefont{D.}~\bibnamefont{Leykam}},
  \bibinfo{author}{\bibfnamefont{A.}~\bibnamefont{Andreanov}},
  \bibnamefont{and} \bibinfo{author}{\bibfnamefont{S.}~\bibnamefont{Flach}},
  \bibinfo{journal}{Adv.\ Phys.:\ X} \textbf{\bibinfo{volume}{3}},
  \bibinfo{pages}{1473052} (\bibinfo{year}{2018}).

\bibitem[{\citenamefont{Andrei et~al.}(2012)\citenamefont{Andrei, Li, and
  Du}}]{ALD12}
\bibinfo{author}{\bibfnamefont{E.~Y.} \bibnamefont{Andrei}},
  \bibinfo{author}{\bibfnamefont{G.}~\bibnamefont{Li}}, \bibnamefont{and}
  \bibinfo{author}{\bibfnamefont{X.}~\bibnamefont{Du}},
  \bibinfo{journal}{Reports on Progress in Physics}
  \textbf{\bibinfo{volume}{75}}, \bibinfo{pages}{056501}
  (\bibinfo{year}{2012}).

\bibitem[{\citenamefont{Senkovskiy
  et~al.}(2018{\natexlab{b}})\citenamefont{Senkovskiy, Usachov, Fedorov,
  Marangoni, Haberer, Tresca, Profeta, Caciuc, Tsukamoto, Atodiresei
  et~al.}}]{VB_B-7AGNR2}
\bibinfo{author}{\bibfnamefont{B.~V.} \bibnamefont{Senkovskiy}},
  \bibinfo{author}{\bibfnamefont{D.~Y.} \bibnamefont{Usachov}},
  \bibinfo{author}{\bibfnamefont{A.~V.} \bibnamefont{Fedorov}},
  \bibinfo{author}{\bibfnamefont{T.}~\bibnamefont{Marangoni}},
  \bibinfo{author}{\bibfnamefont{D.}~\bibnamefont{Haberer}},
  \bibinfo{author}{\bibfnamefont{C.}~\bibnamefont{Tresca}},
  \bibinfo{author}{\bibfnamefont{G.}~\bibnamefont{Profeta}},
  \bibinfo{author}{\bibfnamefont{V.}~\bibnamefont{Caciuc}},
  \bibinfo{author}{\bibfnamefont{S.}~\bibnamefont{Tsukamoto}},
  \bibinfo{author}{\bibfnamefont{N.}~\bibnamefont{Atodiresei}},
  \bibnamefont{et~al.}, \bibinfo{journal}{ACS\ Nano}
 (\bibinfo{year}{in press}{\natexlab{b}}).
\end{thebibliography}
\end{document}